\documentclass[aps, prl, twocolumn, twopage, 10pt, floatfix, longbibliography, nofootinbib, nobibnotes]{revtex4-2}
\usepackage[OT2,T1]{fontenc}

\usepackage{amsmath,amsfonts,amssymb, amsthm}

\usepackage{bm}
\usepackage{graphicx}

\usepackage{multirow}

\usepackage[a4paper, bindingoffset=0cm, left=2cm, right=2cm, top=2cm, bottom=2cm, headsep=0.5cm, footskip=1cm]{geometry}

\usepackage{parskip}
\usepackage{tikz-cd}
\DeclareMathAlphabet{\mathscrbf}{OMS}{mdugm}{b}{n}
\usepackage{tikz}
\usetikzlibrary{decorations.markings}

\usepackage{enumerate}
\usepackage{enumitem}
\usepackage[colorlinks, allcolors=blue]{hyperref}

\usepackage{empheq}

\usepackage[user,titleref]{zref}

\usepackage{tikz}
\usetikzlibrary{shapes.misc, shapes.geometric}

\usepackage{fancyhdr}
\fancypagestyle{plain}{
  \fancyhf{}
  \fancyfoot[C]{\thepage}
  
}

\usepackage{titlesec}
\usepackage[titles]{tocloft}
\cftsetindents{section}{0em}{1.7em}
\cftsetindents{subsection}{1.7em}{1.2em}
\makeatletter
\renewcommand{\paragraph}{%
  \@startsection{paragraph}{4}%
  {\z@}{1.75ex \@plus 1ex \@minus .2ex}{-1em}%
  {\normalfont\normalsize\bfseries}%
}
\makeatother

\newcommand{\Tr}{\operatorname{Tr}}
\newcommand{\dd}{\textrm{d}}
\newcommand{\ee}{\textrm{e}}
\newcommand{\ii}{\textrm{i}}

\def\sgn{\operatorname{sgn}}

\newcommand{\pdag}{^{\vphantom{\dagger}}}

\newcommand{\bjmath}{\boldsymbol{\jmath}}

\newcommand{\bp}{\mathbf{p}}
\newcommand{\bq}{\mathbf{q}}

\newcommand{\bx}{\mathbf{x}}
\newcommand{\by}{\mathbf{y}}

\newcommand{\bJ}{\mathbf{J}}

\newcommand{\bP}{\mathbf{P}}

\newcommand{\bX}{\mathbf{X}}
\newcommand{\bY}{\mathbf{Y}}

\newcommand{\cA}{{\cal A}}
\newcommand{\cB}{{\cal B}}

\newcommand{\cD}{{\cal D}}

\newcommand{\cF}{{\cal F}}
\newcommand{\cG}{{\cal G}}
\newcommand{\cH}{{\cal H}}

\newcommand{\cL}{{\cal L}}
\newcommand{\cM}{{\cal M}}
\newcommand{\cN}{{\cal N}}
\newcommand{\cO}{{\cal O}}
\newcommand{\cP}{{\cal P}}
\newcommand{\cQ}{{\cal Q}}

\newcommand{\cT}{{\cal T}}

\newcommand{\cV}{{\cal V}}

\renewcommand{\Re}{\operatorname{Re}}
\renewcommand{\Im}{\operatorname{Im}}

\newcommand{\ran}{\operatorname{ran}}

\newcommand{\XO}{X}
\newcommand{\KXO}{Y}

\newcommand{\prho}{\lambda}

\begin{document}
%==========================================================

\setlength{\skip\footins}{20pt plus 2pt minus 2pt}

%==========================================================

\title{%
Fluctuation-dissipation structure of quantum geometry%
}%

\author{\vspace{-3mm}Per Moosavi}
\email{pmoosavi@phys.ethz.ch}
\affiliation{Institute for Theoretical Physics, ETH Zurich, Wolfgang-Pauli-Strasse 27, 8093 Z\"urich, Switzerland\looseness=-1\vspace{-1.5mm}}

\date{October 29, 2026}

\begin{abstract}
Geometry of quantum states is a fundamental research area with applications ranging from topological materials in condensed matter to variational algorithms in quantum information. Due to their relative simplicity, pure states are often studied, while mixed states are generally needed, e.g., to have nonzero temperature. Mixed-state geometry, however, is much more challenging and has so far lacked any structure relating distance and curvature, like the K\"ahler structure does for pure states. This work introduces that structure. We show that it has the physical meaning of fluctuation-dissipation---by identifying distance with fluctuations and curvature with linear response---and that it becomes the known structure for pure states, but is non-K\"ahler otherwise. This paints a simple picture that unifies fluctuation-dissipation and geometry of quantum states. It also applies on parameter spaces, as encountered in condensed matter or estimation theory. Our picture implies that the geometry contains all linear responses and fluctuations, proves the fluctuation-dissipation relation, explains that quantum geometry appears in any sum rule, and allows one to geometrically characterize generic transport behaviors. Closed and open systems are treated with equal ease.
\end{abstract}

%==========================================================
\maketitle
%==========================================================

\thispagestyle{fancy}
\pagestyle{fancy}

%==========================================================
\noindent
\csname phantomsection\endcsname%
\addcontentsline{toc}{section}{Introduction}%
%==========================================================
%
The fluctuation-dissipation relation is central to classical and quantum statistical physics \cite{Kubo:1966, MPRV:2008}.
One reason is that it allows one to predict fluctuations from response functions like conductivities or susceptibilities, and vice versa.
Typical examples are charge, heat, or polarization fluctuations versus electrical, thermal, or optical conductivity.
As a subject, it has a long history, dating back to Einstein's theory of Brownian motion \cite{Einstein:1905}, with pioneering work on electrical resistors \cite{Nyquist:1928}, the regression hypothesis \cite{Onsager:1931a}, and nonequilibrium systems \cite{CallenWelton:1951, Kubo:1957}.
Much more recently, there has been active interest in relations between different fluctuations or responses and the geometry of quantum states \cite{SWM:2000, ShitaraUeda:2016, HHTZ:2016, CSV:2018, LVSC:2019, LYLW:2020, CVS:2020, AhnEtAl:2022, LambertSorensen:2023, YuEtAl:2025, VermaQueiroz:2025, JiEtAl:2025, GuanBradlyn:2026, WangEtAl:2026, BittnerAcuna:2026}.
However, a simple clear picture has yet to emerge.

\begin{figure}[b]
\centering

\vspace{-3mm}
\includegraphics[scale=0.07, clip=true, trim=0mm 0mm 0mm 0mm]{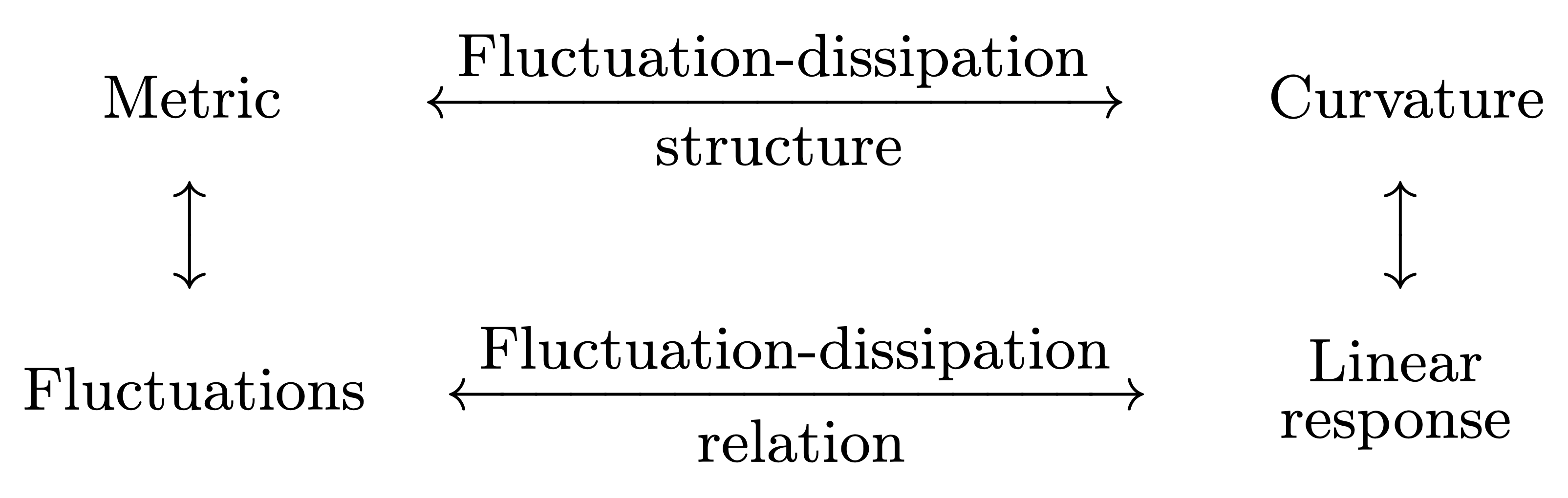}
\vspace{-3mm}

\caption{%
Fluctuation-dissipation and geometry of quantum states.
The top relation [Eq.~\eqref{FDS}] is geometric and holds on the full space of density operators.
The vertical ones [Eqs.~\eqref{kappa_cF}--\eqref{fluct_cG}] are identifications for stationary states under some dynamics (unitary or open).
The bottom one (the fluctuation-dissipation relation) follows from the others.%
}
\label{Fig:Connections}
\vspace{-1mm}
\end{figure}

This paper puts forward such a picture (Fig.~\ref{Fig:Connections}), which unifies two seemingly separate topics:
We claim that fluctuation-dissipation is inherent to the geometry of quantum states, by showing that it is the physical meaning of a structure that relates distance and curvature for all density operators, i.e., mixed states.

\begin{figure}[t]
\centering

\includegraphics[scale=0.068, clip=true, trim=0mm 0mm 0mm 0mm]{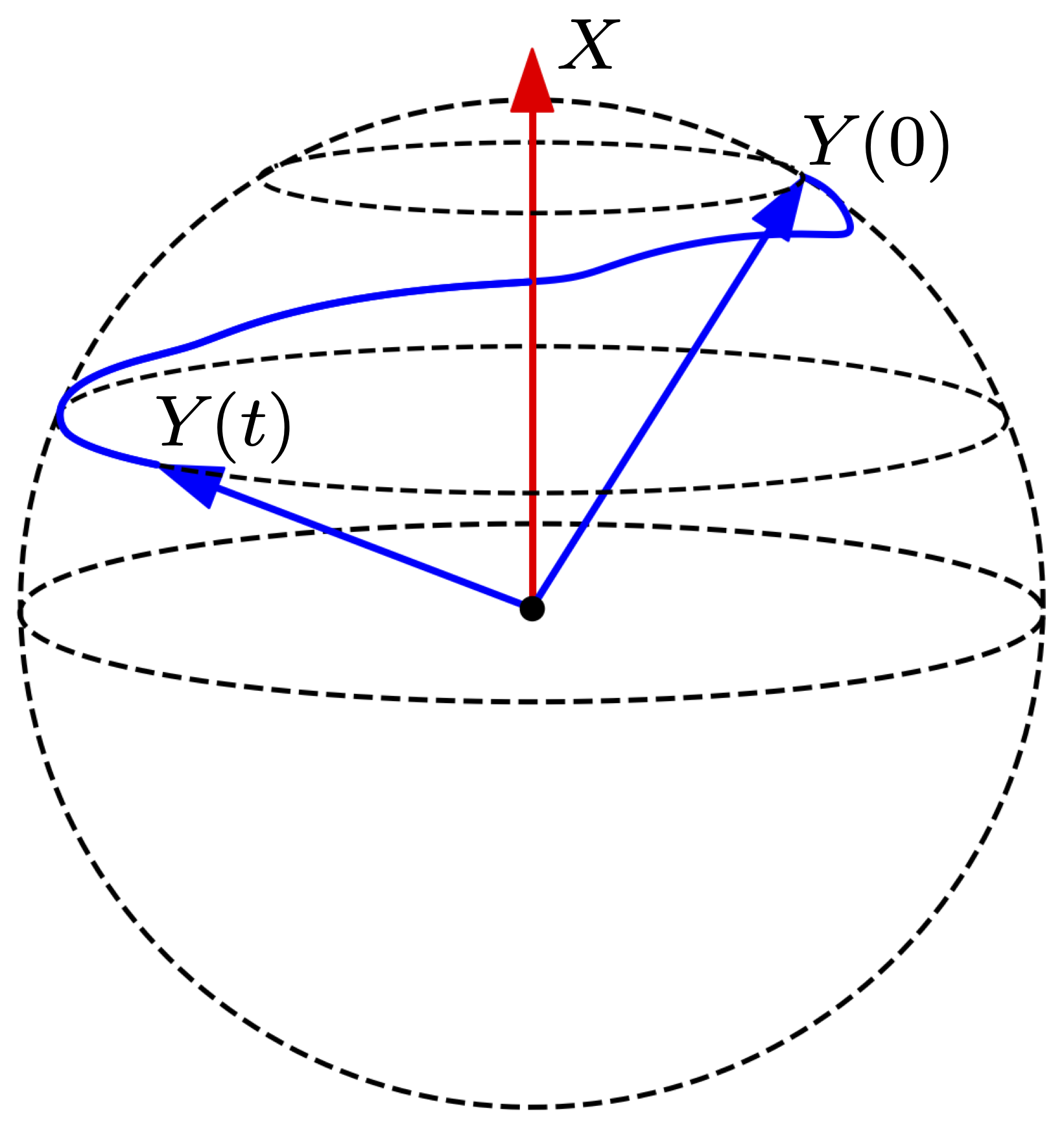}

\vspace{-2.9mm}

\caption{%
Transport in terms of tangent vectors (drawn as 3D analogues) to the space of quantum states:
The vector $\KXO(t)$ evolves in time $t$ around a fixed vector $\XO$, their inner product equal to a time-dependent conductivity [Eq.~\eqref{ashaped_eq}].
For a closed system (case shown), $Y(t)$'s length is constant and only its angle with $\XO$ can vary.
On physical grounds, for the response in a total current [Eq.~\eqref{kappa_gen_grounds} in \hyperref[Sec:Methods]{\ztitleref{Sec:Methods}}], $\KXO(t)$ starts on a small circle, orthogonal to $\XO$, and rotates asymptotically onto a larger parallel circle.
Transport is purely diffusive if the angle goes from $<90^\circ$ to $90^\circ$, and purely ballistic if it is a constant $< 90^\circ$.
For open systems, $\KXO(t)$'s length can shrink and the angle can change.%
}
\label{Fig:TransportSketch}
\vspace{-3mm}
\end{figure}

The upshot is that \emph{quantum fluctuation-dissipation is geometry of quantum states}.
Indeed, from its role in our picture, the structure proves the fluctuation-dissipation relation for all density operators stationary under some added dynamics, regardless if unitary or open, while the geometry itself is independent of such additions.
Among its uses, the picture transfers to parameter spaces---the setting called \emph{quantum geometry} in condensed matter \cite{YuEtAl:2025}---and allows one to geometrically characterize transport behaviors (Fig.~\ref{Fig:TransportSketch}).
The picture (Fig.~\ref{Fig:Connections}) also explains why quantum geometry appears in sum rules \cite{Kubo:1966}, such as in \cite{SWM:2000, HHTZ:2016}, which we derive in their most general forms using our structure.

%==========================================================
\vspace{0.0mm}\noindent\textbf{\fontsize{11}{13.5}\selectfont What is the structure?}\vspace{0.5mm}\\
\csname phantomsection\endcsname%
\addcontentsline{toc}{section}{What is the structure?}%
%==========================================================
%
To begin, the space of quantum states can be equipped with a metric and a curvature.
Simply put, they are instruments for measuring distances between states and rotations of ``arrows'' carried around a loop.
This is generally true for mixed states, but most widely known for pure states.
For the latter, there is a deep connection between what are then called the \emph{Fubini-Study metric} and \emph{symplectic form} through a complex structure, i.e., the geometry is \emph{K\"ahler}.
Things are more intricate for mixed states, with infinitely many metrics meeting the desired criteria (to be monotone and Riemannian) \cite{BeZy:2009}.
That said, there is arguably a natural pair---the \emph{Bures metric} and the \emph{Uhlmann curvature} \cite{Uhlmann:1986, Uhlmann:1991}---that reduce to the pure case,\footnote{E.g., the Kubo-Mori metric, arising from relative entropy \cite{BaALRe:1986}, instead diverges for pure states \cite{BeZy:2009}.
In fact, for finite-dimensional Hilbert spaces, all monotone Riemannian metrics are classified \cite{BeZy:2009}, and the Bures metric is the unique one that is both Fisher-Rao and Fubini-Study adjusted \cite{PetzSudar:1996}.} but any structure relating the two has yet to be appreciated.

We introduce this structure (Definition~\hyperref[Def:0]{\ztitleref{Def:0}} and Theorem~\hyperref[Theorem:1]{\ztitleref{Theorem:1}}) and prove that it is K\"ahler for pure states only (Theorem~\hyperref[Theorem:2]{\ztitleref{Theorem:2}}).
The main claim is that its physical meaning is \emph{fluctuation-dissipation}.
To make this point, we show that the structure yields the expected fluctuation-dissipation form for thermal states \cite{CallenWelton:1951, Kubo:1957, Kubo:1966}:
At inverse temperature $\beta $, the metric equals the curvature multiplied by the distribution function $\frac{1}{2} \coth(\beta \omega_{jk}/2)$ with energy differences $\omega_{jk}$
[Eq.~\eqref{Krho_beta_jk} in \hyperref[Sec:Methods]{\ztitleref{Sec:Methods}}].\footnote{We use units so that $\hbar = 1$.}
While suggestive, the claim is not mere similarity in appearance:
We show that the Bures metric and the Uhlmann curvature are synonymous with fluctuations and linear responses for stationary density operators.
Our structure then proves the usual relation between the latter two for closed systems, and even extends it to open systems (cf.\ Theorem~\hyperref[Theorem:3]{\ztitleref{Theorem:3}}).
This motivates the structure's name.
(\emph{Fluctuation-response} is more accurate, since it includes nondissipative responses, but we use the more common name, as does \cite{Kubo:1966}.)
Our results also endow the K\"ahler structure
for pure states with the fluctuation-dissipation meaning, while noting that K\"ahler is irrelevant for its use here.

The picture in Fig.~\ref{Fig:Connections} covers important results that have appeared on parameter spaces for different physical systems.
For pure states, this includes the Souza-Wilkens-Martin (SWM) sum rule \cite{SWM:2000} for many-body electronic systems, which relates optical conductivity with the metric via polarization fluctuations.
We show that our picture explains this rule by deriving its most general form, without using many-body physics, and it also explains the observation \cite{AhnEtAl:2022} that optical responses are geometric in nature.
Another instance comes from metric contributions to the superfluid weight, e.g., for flat-band superconductors \cite{PeottaTorma:2015}, discussed in \cite{YuEtAl:2025} along with other examples.
We point out that all its contributions have a geometric origin due to our picture.
For mixed states, we mention work \cite{JiEtAl:2025} on time-dependent metric and curvature for (closed) many-body systems at nonzero temperature, motivated by \cite{VermaQueiroz:2025} and SWM-type sum rules for thermal states \cite{HHTZ:2016}.
The metric in \cite{JiEtAl:2025} is based on quantum Fisher information \cite{BraunsteinCaves:1994}, which originates from estimation theory, and corresponds to the Bures metric \emph{pulled back} by a map to some parameter space.
For comparison, our geometric quantities and relations are on the full space of density operators, independent of dynamics.
This \emph{universality} is key to why our picture applies
to general quantum systems, closed or open,
with the minimal requirement that there is a stationary density operator.
Further examples of our results will appear in \cite{MBLO:2026} for emergent field-theory descriptions of quantum many-body systems, where parameter space consists of conformal transformations \cite{LMO:2025}.

%==========================================================
\vspace{0.0mm}\noindent\textbf{\fontsize{11}{13.5}\selectfont Geometry of quantum states}\vspace{0.5mm}\\
\csname phantomsection\endcsname%
\addcontentsline{toc}{section}{Geometry of quantum states}%
%==========================================================
%
Let $\cH$ be a Hilbert space and $\cD$ the space of \emph{density operators} on $\cH$, i.e., Hermitian nonnegative operators $\rho$ with trace $\Tr \rho = 1$.\footnote{\label{Note:Separable}$\cH$ need not be finite dimensional, but generally a \emph{separable} (infinite-dimensional) Hilbert space; see Part~\hyperref[Part:A1]{\ztitleref{Part:A1}} in \hyperref[Sec:Methods]{\ztitleref{Sec:Methods}}.}
This $\cD$ is what we here mean by the space of (\emph{mixed}) \emph{quantum states}.
Its interior consists of all positive operators, while operators on the boundary $\partial\cD$ have zero as an eigenvalue.
Pure states are special cases on $\partial\cD$ with exactly one nonzero eigenvalue and can be written as $\rho_\psi = |\psi\rangle\langle\psi|/\langle\psi|\psi\rangle$ for some nonzero $|\psi\rangle$ in $\cH$.

We consider the Bures metric $\cG$ and the (scalar) Uhlmann curvature $\cF$ given by
\begin{subequations}
\label{cGcF_XY}
\begin{align}
\cG_{\rho}(X,Y)
& = \frac{1}{2} \Tr \rho \, \bigl\{ G_{X}, G_{Y} \bigr\}, \label{cG_XY} \\
\cF_{\rho}(X,Y)
& = -\ii \Tr \rho \, \bigl[ G_{X}, G_{Y} \bigr] \label{cF_XY}
\end{align}
\end{subequations}
for tangent vectors, i.e., traceless Hermitian operators, $X$, $Y$ at $\rho$ in $\cD$.\footnote{\label{Note:Tangent_vectors}All possible $X$, $Y$ are allowed at interior points of $\cD$, while at points on $\partial\cD$, they are restricted to not point out of $\cD$; see Part~\hyperref[Part:A3]{\ztitleref{Part:A3}} in \hyperref[Sec:Methods]{\ztitleref{Sec:Methods}} also for when $\dim(\cH) = \infty$.}
Here, $\cG$ and $\cF$ are written as the anticommutator $\{ \cdot, \cdot \}$ and commutator $[ \cdot, \cdot ]$ of Hermitian operators $G_X$, $G_Y$ that solve the Lyapunov equation\footnote{\label{Note:GUniqueness}The solution is unique for $\rho$ in the interior of $\cD$. On $\partial\cD$, it is only determined along directions that do not point out of $\cD$, but those are also the only ones that contribute in Eq.~\eqref{cGcF_XY}.}
\begin{equation}
\label{Lyapunov_eq}
G_{X} \rho + \rho\hspace{0.5pt} G_{X} = X,
\end{equation}
and similarly for $G_{Y}$.
Eq.~\eqref{Lyapunov_eq} is key: It links $\cD$ with the bundle of quantum state purifications from which Eqs.~\eqref{cG_XY}--\eqref{cF_XY} originate \cite{BeZy:2009, Uhlmann:1991} (see Part~\hyperref[Part:A2]{\ztitleref{Part:A2}} in \hyperref[Sec:Methods]{\ztitleref{Sec:Methods}}).
For pure states, one does not encounter it, since no purification is needed, reflected in that $G_X = X$ can be taken as the solution, with Eq.~\eqref{cGcF_XY} reducing to the Fubini-Study metric and symplectic form [Eq.~\eqref{cGcF_XY_pure_jk} in \hyperref[Sec:Methods]{\ztitleref{Sec:Methods}}].
In quantum geometry, the pure-state metric-curvature pair on parameter space is known as the \emph{quantum metric} \cite{ProvostVallee:1980} and \emph{Berry curvature} \cite{Berry:1984, Simon:1983}.

One can combine $\cG$ and $\cF$ into the \emph{Bures-Uhlmann} or ``\emph{mixed-state quantum geometric}''
\emph{tensor}
\begin{equation}
\label{cQ_XY}
\cQ_{\rho}(X,Y)
\equiv \cG_{\rho} + \frac{\ii}{2} \hspace{0.5pt} \cF_{\rho}
= \Tr \rho \, G_{X} G_{Y},
\end{equation}
just as in the better-known pure case \cite{ProvostVallee:1980}.
I.e., $\cG$ is the symmetric or real part, $\cG = \Re \cQ$, and $\cF/2$ is the antisymmetric or imaginary part, $\cF/2 = \Im \cQ$.
More important to us is the structure that relates $\cG$ and $\cF$:

\noindent
{\bf Definition 0.}
\csname phantomsection\endcsname%
\label{Def:0}%
\makeatletter\def\zref@titleref@current{0}\makeatother
\zlabel{Def:0}%
Introduce $K_{\rho}$ as the map\footnote{I.e., $K_{\rho}$ is a $(1,1)$ tensor, while $\cG_{\rho}$ and $\cF_{\rho}$ are $(0,2)$ tensors.}
between tangent vectors to $\cD$ at $\rho$ that satisfies
\begin{equation}
\label{FDS}
2\cG_{\rho}(K_{\rho} X, Y) = \cF_{\rho}(X,Y)
\end{equation}
for all allowed $X$ and $Y$.
The \emph{fluctuation-dissipation structure} is the family of these maps for all $\rho$ in $\cD$.

Since $\cG_{\rho}$ is nondegenerate (just as for
any Riemannian metric), Eq.~\eqref{FDS} determines the action of $K_{\rho}$ on $X$.
While such an object is not new in itself, it has not been introduced for the Bures-Uhlmann pair of metric and curvature.
We prove (see Part~\hyperref[Part:A5]{\ztitleref{Part:A5}} in \hyperref[Sec:Methods]{\ztitleref{Sec:Methods}}):

\noindent
{\bf Theorem 1.}
\csname phantomsection\endcsname%
\label{Theorem:1}%
\makeatletter\def\zref@titleref@current{1}\makeatother
\zlabel{Theorem:1}%
$K_\rho X$ in Eq.~\eqref{FDS} is the solution to\footnote{Eq.~\eqref{Riesz_rep_eq} determines $K_\rho X$ to the extent Eq.~\eqref{Lyapunov_eq} does for $G_X$.}
\begin{equation}
\label{Riesz_rep_eq}
(K_{\rho} X) \rho + \rho (K_{\rho} X)
= \ii \bigl[ X, \rho \bigr].
\end{equation}

Eqs.~\eqref{cGcF_XY}--\eqref{Riesz_rep_eq} in terms of matrices give valuable insights and make sense
both for $\dim(\cH) < \infty$ and $= \infty$ (see Part~\hyperref[Part:A4]{\ztitleref{Part:A4}} in \hyperref[Sec:Methods]{\ztitleref{Sec:Methods}}).
That said, using a basis is not necessary, so we will continue without one here.

Crucially, $K_\rho$ is defined whether or not $\cF_{\rho}$ is nondegenerate, which is not guaranteed (i.e., it is not necessarily a symplectic form).
This is different from pure $\rho$, and reflects that $K_\rho$ need not be invertible,
let alone satisfy $K_\rho^2 X = -X$, as would be necessary for it to be K\"ahler.
We prove (see again Part~\hyperref[Part:A5]{\ztitleref{Part:A5}} in \hyperref[Sec:Methods]{\ztitleref{Sec:Methods}}):

\noindent
{\bf Theorem 2.}
\csname phantomsection\endcsname%
\label{Theorem:2}%
\makeatletter\def\zref@titleref@current{2}\makeatother
\zlabel{Theorem:2}%
$K_{\rho}$ in Eq.~\eqref{FDS} is a complex structure, thus the geometry is K\"ahler, if and only if $\rho$ is pure.

That $\cF_\rho$ may be degenerate and $K_\rho$ noninvertible are consistent with the divergence of $\frac{1}{2} \coth(\beta \omega_{jk}/2)$ when $\omega_{jk} = 0$ (or $\beta \to 0$) in the fluctuation-dissipation relation for thermal states.
As we show below, Eq.~\eqref{FDS} carries precisely that meaning, justifying the name.

%==========================================================
\vspace{0.0mm}\noindent\textbf{\fontsize{11}{13.5}\selectfont Fluctuation-dissipation meaning}\vspace{0.5mm}\\
\csname phantomsection\endcsname%
\addcontentsline{toc}{section}{Fluctuation-dissipation meaning}%
%==========================================================
%
We start with linear response for closed systems \cite{Kubo:1957}.
Consider a quantum system with Hamiltonian $H$ initially in a stationary state $\rho$, i.e., $[\rho, H] = 0$.\footnote{$H$ must have at least one eigenstate so that at least one $\rho$ in $\cD$ is a stationary state. That $\rho$ is generally not thermal, nor even in equilibrium as it may carry nonzero currents.}
Let $\cO_a$ be observables (Hermitian operators) with labels $a$ and zero average, $\Tr \rho \, \cO_a = 0$.
(The latter can always be imposed by replacing $\cO_a$ by $\cO_a - \Tr \rho \, \cO_a$.)
We measure the linear response $\kappa_{ab}(t)$ in the expectation value of some $\cO_a$ with respect to $\rho$ evolved in time $t > 0$ under the Hamiltonian perturbed by $\cO_b$.
Similar to \cite{Kubo:1966}, one obtains (see Part~\hyperref[Part:B1]{\ztitleref{Part:B1}} in \hyperref[Sec:Methods]{\ztitleref{Sec:Methods}}):
\begin{equation}
\label{kappa_ab_result}
\kappa_{ab}(t)
= -\ii \Tr \rho \bigl[ \cO_a(t), \cO_b \bigr] \qquad (t > 0)
\end{equation}
for time-evolved $\cO_a(t) \equiv \ee^{\ii Ht} \cO_a \ee^{-\ii Ht}$.
It covers both direct and alternating responses, seen from the Fourier transform $\chi_{ab}(\omega) = \int_0^{\infty} \dd t\, \kappa_{ab}(t) \ee^{-\ii\omega t}$ to frequency space, with the direct response being its $\omega = 0$ value.

Eq.~\eqref{kappa_ab_result} by itself resembles the Uhlmann curvature \eqref{cF_XY}.
To show this, we reverse the logic of Eq.~\eqref{Lyapunov_eq}: Instead of $X$ given and $G_X$ unknown, we use it to define tangent vectors from the observables $\cO_a(t)$.
Indeed, since $\cO_a(t)^\dagger = \cO_a(t)$ and $\Tr \rho \, \cO_a(t) = 0$, each of
\begin{subequations}
\label{XaYa_t}
\begin{align}
\XO_a(t)
& \equiv \cO_a(t) \rho + \rho \cO_a(t),
\qquad
\XO_a \equiv \XO_a(0), \label{Xa_t} \\
\KXO_a(t)
& \equiv K_\rho \XO_a(t) \label{Ya_t}
\end{align}
\end{subequations}
with $K_\rho$ in Eqs.~\eqref{FDS}--\eqref{Riesz_rep_eq}
is a tangent vector at $\rho$.${}^{\text{\ref{Note:Tangent_vectors}}}$
Thus,
\begin{equation}
\label{kappa_cF}
\kappa_{ab}(t) = \cF_{\rho}(\XO_a(t), \XO_b)
\qquad (t > 0).
\end{equation}
In other words, the linear response $\kappa_{ab}(t)$ is precisely the Uhlmann curvature \eqref{cF_XY} for an evolving tangent vector $\XO_a(t)$ and a fixed one $\XO_b$ defined by Eq.~\eqref{Xa_t}.
While similar formulas have appeared for pure states, e.g., to compute the Hall conductivity from the Berry curvature \cite{TKNN:1982}, we are not aware of any general identification like Eq.~\eqref{kappa_cF} for all stationary density operators.

We now turn to fluctuations.
They are given by the Bures metric for the same tangent vectors \eqref{Xa_t}:
\begin{equation}
\label{fluct_cG}
\cG_{\rho}(\XO_a(t), \XO_b)
= \frac{1}{2} \Tr \rho \, \{ \cO_a(t), \cO_b \}.
\end{equation}
These are symmetrized correlations for our observables $\cO_a$---the typical meaning of fluctuations in quantum statistical physics \cite{Kubo:1957}.
Relatedly, the corresponding tensor \eqref{cQ_XY} is a covariance matrix, $\cQ_{\rho}(\XO_a(t), \XO_b) = \Tr \rho \, \cO_a(t) \cO_b \equiv \operatorname{Cov}_{\rho}(\cO_a(t), \cO_b)$, again since the averages $\Tr \rho \, \cO_a(t) = 0$, and the fluctuations are simply $\cG_{\rho}(\XO_a(t), \XO_b) = \Re \operatorname{Cov}_{\rho}(\cO_a(t), \cO_b)$.
Since $G_X = X$ for pure $\rho$, this bestows the fluctuation interpretation on the Fubini-Study metric [Eq.~\eqref{PV_eq} in \hyperref[Sec:Methods]{\ztitleref{Sec:Methods}}], as observed in \cite{ProvostVallee:1980} for
the quantum metric.
Our identification \eqref{fluct_cG} covers both statistical (e.g., thermal) and quantum fluctuations, as opposed to only quantum ones for pure states.
Indeed, the Bures metric splits into classical and quantum parts [Eq.~\eqref{cG_XY_jk_split} in \hyperref[Sec:Methods]{\ztitleref{Sec:Methods}}] \cite{BraunsteinCaves:1994}.
Due to Eq.~\eqref{FDS}, the curvature must similarly split.
Since all diagonal elements of $K_\rho X$ are zero [Eq.~\eqref{Krho_jk} in \hyperref[Sec:Methods]{\ztitleref{Sec:Methods}}], only the quantum part of the metric survives in Eq.~\eqref{FDS}, consistent with that the linear response \eqref{kappa_ab_result} and the curvature \eqref{kappa_cF} are quantum.

The tangent vectors $\XO_a$ in Eq.~\eqref{Xa_t} are general: They ordinarily derive not from curves given by unitary flows,\footnote{As for a coadjoint orbit \cite{Kirillov:2004, Heydari:2015}, whose tangent space formally corresponds to a subfamily of the full tangent space of $\cD$.} but trace-preserving completely positive maps \cite{Lindblad:1976}.
Still, for closed systems, their evolution is $\XO_a(t) = \ee^{\ii Ht} \XO_a \ee^{-\ii Ht}$ for $t > 0$, i.e., a unitary rotation of $\XO_a$.
Their lengths $\ell_{\rho}[\XO_a(t)] \equiv \sqrt{\cG_{\rho}( \XO_a(t), \XO_a(t))}$ measured by the Bures metric \eqref{cG_XY} are thus constant in time, as are the lengths $\ell_{\rho}[\KXO_a(t)]$ of $\KXO_a(t)$ in Eq.~\eqref{Ya_t} since $\KXO_a(t) = \ee^{\ii Ht} K_\rho \XO_a \ee^{-\ii Ht}$ due to $[K_{\rho}, H] = 0$.

The identifications \eqref{kappa_cF}--\eqref{fluct_cG} do not require unitary dynamics, just $\cO_a(t)$ Hermitian and $\Tr \rho \cO_a(t) = 0$ for all $t \geq 0$.${}^{\text{\ref{Note:Tangent_vectors}}}$
We can thus extend to Hamiltonian perturbations of open systems (see Part~\hyperref[Part:B2]{\ztitleref{Part:B2}} in \hyperref[Sec:Methods]{\ztitleref{Sec:Methods}}).
While more restrictive than necessary, we consider the Markovian case of Lindblad dynamics $\Phi_t = \ee^{t \cL}$ generated by $\cL \rho = -\ii[H , \rho] + \cL_{\text{D}} \rho$ with dissipator $\cL_{\text{D}}$ \cite{Lindblad:1976}.
For a system in an initial stationary state $\rho$, i.e., $\cL \rho = 0$, Eqs.~\eqref{kappa_ab_result}--\eqref{fluct_cG} are the same, just with $\cO_a(t) \equiv \ee^{t \cL^*} \cO_a$ using the adjoint Liouvillian $\cL^* = \ii[H , \bullet] + \cL_{\text{D}}^*$.
This means that the tangent vectors $X_a(t)$ and $Y_a(t)$ do not rotate unitarily.
Instead, their lengths can shrink according to $\ell_{\rho}[\KXO_a(t)] \leq \ell_{\rho}[\XO_a(t)] \leq \ell_{\rho}[\XO_a(s)]$ for $t \geq s \geq 0$.

%==========================================================
\vspace{0.0mm}\noindent\textbf{\fontsize{11}{13.5}\selectfont The picture and its consequences}\vspace{0.5mm}\\
\csname phantomsection\endcsname%
\addcontentsline{toc}{section}{The picture and its consequences}%
%==========================================================
%
The complete picture is given in Fig.~\ref{Fig:Connections}.
The top relation is Eq.~\eqref{FDS}, which holds for all  $\rho$ in $\cD$.
Eqs.~\eqref{kappa_ab_result} and \eqref{kappa_cF} form the right relation, and Eq.~\eqref{fluct_cG} is the left one, established for all $\rho$ stationary under an added dynamics.
Crucially, only the tangent vectors \eqref{XaYa_t} depend on time, while the geometry does not.
Together, these relations prove the bottom one---the fluctuation-dissipation relation---for any stationary $\rho$ for closed and open systems alike.
This shows that the structure \eqref{FDS} is more fundamental than the usual relation.

The above (together with Part~\hyperref[Part:A3]{\ztitleref{Part:A3}} in \hyperref[Sec:Methods]{\ztitleref{Sec:Methods}}) proves that the whole picture (Fig.~\ref{Fig:Connections}) for a given dynamics condenses into a single \emph{a-shaped} equation:

\noindent
{\bf Theorem 3.}
\csname phantomsection\endcsname%
\label{Theorem:3}%
\makeatletter\def\zref@titleref@current{3}\makeatother
\zlabel{Theorem:3}%
Given a Liouvillian
$\cL = - \ii [H, \bullet] + \cL_{\text{D}}$
with a stationary density operator $\rho$.
Then, for $t > 0$,
\begin{align}
2\cG_{\rho}(\KXO_a(t), \XO_b) \; & = \quad \cF_{\rho}(\XO_a(t), \XO_b) \nonumber \\
\rotatebox{90}{$=$} \qquad\quad & \phantom{=} \qquad\qquad\quad \rotatebox{90}{$=$} \label{ashaped_eq} \\
\Tr \rho \, \{ \cP_a(t), \cO_b \}
& = -\ii \Tr \rho \bigl[ \cO_a(t), \cO_b \bigr]
= \kappa_{ab}(t) \nonumber
\end{align}
with tangent vectors $\XO_a(t)$, $\KXO_a(t)$ in Eq.~\eqref{XaYa_t} and observables $\cO_a(t) = \ee^{t \cL^*} \cO_a$,  $\cP_a(t) \equiv G_{\KXO_a(t)}$ [cf.\ Eq.~\eqref{Lyapunov_eq}].

Our picture (Fig.~\ref{Fig:Connections}) also applies to parameter spaces $\cM$---the setting encountered in condensed matter and estimation theory---where one considers a family of states $\rho_\alpha$ labeled by $\alpha$ in $\cM$, e.g., representing possible physical manipulations.
The only nontrivial criteria are that the metric induced on $\cM$ is nondegenerate (thus Riemannian) and that the parameter space is large enough (see Part~\hyperref[Part:B6]{\ztitleref{Part:B6}} in \hyperref[Sec:Methods]{\ztitleref{Sec:Methods}}).
That said, $\cD$ is different from parameter space in that there is generally no ``good'' parametrization of the whole $\cD$.
However, introducing a parametrization is a priori not necessary, depending on the application.
We make this case by deriving sum rules \cite{Kubo:1966} that involve the geometry:
Take any response function, extract its dissipative part, and compute its moments in frequency space (i.e., $\omega^n$-weighted frequency integrals) to get sum rules that thanks to Eq.~\eqref{ashaped_eq} can always be expressed in geometric terms.
These results can then be pulled back to a parameter space of interest.

Case in point: We derive the general form of the SWM sum rule \cite{SWM:2000}.
For any closed system with a nondegenerate ground state $|\psi\rangle$ (the rest of the spectrum fully general), we obtain
(see Part~\hyperref[Part:B5]{\ztitleref{Part:B5}} in \hyperref[Sec:Methods]{\ztitleref{Sec:Methods}}):\vspace{-1mm}
\begin{equation}
\label{SWM_sum_rule_GS}
\int_0^\infty \dd\omega\, \frac{\Re \chi_{ab}^{\textrm{S}}(\omega)}{\omega}
= \pi \cG_{\rho_{\psi}}(X_a,X_b) \vspace{-1.5mm}
\end{equation}
for the Fourier transform $\chi_{ab}^{\textrm{S}}(\omega)$ of the ``symmetrized optical conductivity'' $\sigma_{ab}^{\textrm{S}}(t) = \frac{\ii}{2} \partial_t \bigl( \Tr \rho_\psi \bigl[ \cO_a(t), \cO_b \bigr] +  \Tr \rho_\psi \bigl[ \cO_b(t), \cO_a \bigr] \bigr)$ for $t > 0$.
In a sense, this relates optical conductivity with the metric ``diagonally across'' the picture in Fig.~\ref{Fig:Connections}.
However, we are not limited to ground states:
General sum rules for any stationary pure or mixed state $\rho$ in $\cD$ are straightforward to derive [Eqs.~\eqref{SWM_sum_rule_statpure}, \eqref{SWM_sum_rule_statpure_scattering}, and \eqref{SWM_sum_rule_statmix} in \hyperref[Sec:Methods]{\ztitleref{Sec:Methods}}], which explain how to extend the main result in \cite{HHTZ:2016} beyond thermal states.
These sum rules are wholly contained within our picture: They are obtained using only our
structure \eqref{FDS} and Eq.~\eqref{ashaped_eq}, adjusted in this case to measure the response in $-\partial_t \cO_a(t)$ instead of $\cO_a(t)$.

Another use of our approach concerns the superfluid weight \cite{PeottaTorma:2015}.
Since it comes from a current-current susceptibility [Eq.~\eqref{jj_response} in \hyperref[Sec:Methods]{\ztitleref{Sec:Methods}}] \cite{YuEtAl:2025}, Eq.~\eqref{ashaped_eq} directly relates it to the metric.
This means that all contributions to the weight are geometric, just as any quantum response function must be from our picture.

Finally, Eq.~\eqref{ashaped_eq} allows one to characterize transport behaviors in simple geometric terms.
Indeed, $\kappa_{ab}(t) = 2\cG_{\rho}(\KXO_a(t), \XO_b)$ says that $\kappa_{ab}(t)$ is the inner product of tangent vectors of the form \eqref{XaYa_t}.
For closed systems, since the length of $\KXO_a(t)$ is constant, this relation says that $\KXO_a(t)$ rotates on a higher-dimensional sphere in the tangent space with the angle relative to the fixed $\XO_b$ given by $\kappa_{ab}(t)$.
On physical grounds, for the response in a total current [Eq.~\eqref{kappa_gen_grounds} in \hyperref[Sec:Methods]{\ztitleref{Sec:Methods}}], the different types of behaviors are:
a) \emph{Zero conductivity}:
$\KXO_a(t)$ lies on the great circle orthogonal to $\XO_b$.
b) \emph{Pure ballistic}:
$\KXO_a(t)$ traces out a cone around $\XO_b$.
c) \emph{Pure diffusive}:
$\KXO_a(t)$ decays onto the great circle orthogonal to $\XO_b$.
The generic behavior is sketched in Fig.~\ref{Fig:TransportSketch}, as a 3D analogue of the tangent space.
For open systems, the inner product is still a conductivity, but the vectors $\KXO_a(t)$ can also decay inside the sphere.

%==========================================================
\vspace{0.0mm}\noindent\textbf{\fontsize{11}{13.5}\selectfont Conclusion and outlook}\vspace{0.5mm}\\
\csname phantomsection\endcsname%
\addcontentsline{toc}{section}{Conclusion and outlook}%
%==========================================================
%
Our work introduced a structure revealing that quantum fluctuation-dissipation is geometry of quantum states.
The implications for quantum physics are evident, as demonstrated in regards to sum rules \cite{SWM:2000, HHTZ:2016}, the superfluid weight \cite{PeottaTorma:2015, YuEtAl:2025},
and equal treatment of closed and open systems.
They may be most immediate in quantum information geometry \cite{BeZy:2009, SidhuKok:2020} and quantum matter \cite{Torma:2023, YuEtAl:2025},
but more broadly consequential anywhere quantum geometry or fluctuation-dissipation arises, including in high-energy theory \cite{ErdmengerGrosvenorJefferson:2020}.

Probing mixed-state geometry is crucial for understanding quantum matter such as topological insulators and superconductors beyond ground-state properties \cite{BudichDiehl:2015, HZWGC:2024}.
The same goes for variational quantum algorithms \cite{CerezoEtAl:2021}, e.g., extending natural gradient descent \cite{StokesEtAl:2020} beyond pure states \cite{MinerviniPatelWilde:2025}.
Our results imply that linear response is such a probe:
As a consequence of our picture (Fig.~\ref{Fig:Connections}), any response function probes the curvature of the space of mixed quantum states, and via the fluctuation-dissipation structure, also its metric.
The converse claim is even sharper: The geometry of mixed states contains all linear responses and fluctuations, for closed and open systems.
These claims may be used in theoretical and experimental contexts alike.
E.g., to measure the Bures metric and the Uhlmann curvature for specific systems, or to estimate them by computing response functions within effective descriptions \cite{Giamarchi:2003, CCGOR:2011, Doyon:2020}.
Our results open the door to use even classical such descriptions, such as hydrodynamics \cite{Spohn:1991}, to extract information on the mixed-state geometry of any described quantum system.

%==========================================================
\vspace{0.0mm}\noindent\textbf{\fontsize{11}{13.5}\selectfont Acknowledgments}\vspace{0.5mm}\\
\csname phantomsection\endcsname%
\addcontentsline{toc}{section}{Acknowledgments}%
%==========================================================
%
I am thankful to Mathieu Beauvillain, Bastien Lapierre, and Blagoje Oblak for collaborations on related work and Alexander Bols, Matteo Delladio, J\"urg Fr\"ohlich, Gian Michele Graf, Ilaria Maccari, Konstantin Merz, and Ivo Souza for helpful remarks and discussions.
I am particularly grateful to Krzysztof Gaw\k{e}dzki for exchanges in 2019--2020 on linear response theory.

%==========================================================

%\bibliography{references_FDSofQG}


%==========================================================
\onecolumngrid
\clearpage
\twocolumngrid
%==========================================================

\noindent\textbf{\fontsize{12}{13.5}\selectfont Methods}\vspace{1mm}
\csname phantomsection\endcsname%
\addcontentsline{toc}{part}{Methods}
\label{Sec:Methods}
\makeatletter\def\zref@titleref@current{Methods}\makeatother
\zlabel{Sec:Methods}%
%==========================================================

\vspace{-1.5mm}

\noindent
The methods consist of two parts:
Part~\hyperref[Part:A]{\ztitleref{Part:A}} on Hilbert space theory, and Part~\hyperref[Part:B]{\ztitleref{Part:B}} on linear response theory.

\vspace{0.5mm}

%==========================================================
\vspace{0.0mm}\noindent\textbf{\fontsize{11}{13.5}\selectfont A. Hilbert space theory}\vspace{0.5mm}
\csname phantomsection\endcsname%
\addcontentsline{toc}{section}{A. Hilbert space theory}%
\label{Part:A}
\makeatletter\def\zref@titleref@current{A}\makeatother
\zlabel{Part:A}%
%==========================================================

\vspace{-2mm}

%==========================================================
\vspace{0.0mm}\noindent\textbf{A.1. Some relevant notions for Hilbert spaces}\vspace{0.5mm}\\
\csname phantomsection\endcsname%
\addcontentsline{toc}{subsection}{A.1. Some relevant notions for Hilbert spaces}%
\label{Part:A1}%
\makeatletter\def\zref@titleref@current{A.1}\makeatother
\zlabel{Part:A1}%
%==========================================================
%
We recall some notions for our Hilbert space $\cH$: \vspace{-1mm}
\begin{enumerate}[label = (\alph*), topsep=0.25ex, itemsep=-0ex, partopsep=0.25ex, parsep=0.75ex, leftmargin=4.00ex]

\item
$\cH$ can be infinite dimensional as long as it is \emph{separable}, i.e., it has a countable orthonormal basis.

\item
If $\cH$ is infinite dimensional, an operator $T$ satisfying $T = T^\dagger$ indicates that it is \emph{self-adjoint} and not merely \emph{Hermitian}.
These notions are the same for bounded operators.

\item
A self-adjoint operator $T$ is said to be nonnegative (positive), denoted as $T \geq 0$ ($> 0$), if $\langle\psi| T |\psi\rangle \geq 0$ ($> 0$) for all nonzero $|\psi\rangle \in \cH$.

\item
By quantum states we only mean density operators, i.e., the space $\cD$ of trace-class operators $\rho$ on $\cH$ so that $\rho = \rho^\dagger$, $\rho \geq 0$, and $\Tr \rho = 1$.
More generally, these would be called \emph{normal states} \cite{BratteliRobinson1:1987}.

\item
Any $\rho \in \cD$ is bounded and compact.

\item
By the spectral theorem for compact self-adjoint operators, every $\rho \in \cD$ can be decomposed as $\rho =\sum_j \prho_j |e_j\rangle\langle e_j|$ with eigenvalues $\prho_j \geq 0$.

\item
Due to the above,
$\rho \in \cD$ having full rank, i.e., trivial kernel $\ker \rho \equiv \{ \hspace{0.5pt} |\psi\rangle \in \cH : \rho|\psi\rangle = 0 \hspace{0.5pt} \} = \{ 0 \}$, is equivalent to $\rho > 0$, which is equivalent to all eigenvalues $\prho_j > 0$.

\item
Conversely, $\rho \in \cD$ having nonfull rank is equivalent to $\rho$ having zero as an eigenvalue.

\item
Separability of $\cH$ ensures that the space of full-rank $\rho$ is not empty.
E.g., $\rho = \sum_{j = 1}^{\infty} 2^{-j} |e_j\rangle\langle e_j|$ is in there since $\prho_j = 2^{-j} > 0$ and $\sum_{j=1}^{\infty} 2^{-j} = 1$.

\item
Dividing $\cD$ into full- versus nonfull-rank operators is always possible.
However, viewing the former as the interior of $\cD$ and the latter as the boundary $\partial\cD$ is with respect to some suitable topology.
This is only an issue if $\dim(\cH) = \infty$, in which case ``interior'' and ``boundary'' may alternatively here be taken to be mere labels for the full- and nonfull-rank cases.

\item
If $\dim(\cH) = \infty$, the allowed tangent vectors to $\cD$ are restricted as in Part~\hyperref[Part:A3]{\ztitleref{Part:A3}} below.

\item
Any dynamics is required to have at least one stationary $\rho \in \cD$.
Given a dynamical map $\Phi_t$, this means that there is at least one $\rho$ such that $\Phi_t \rho = \rho$ for all $t$.
For a Markovian open system (without explicit time dependence), $\Phi_t = \ee^{t \cL}$ in terms of a Liouvillian $\cL$, so the set $\{ \hspace{0.5pt} \rho \in \cD : \cL\rho = 0 \hspace{0.5pt} \}$ must be nonempty.
For a closed system, $\cL = -\ii[H, \bullet]$ in terms of a Hamiltonian $H$, and the requirement translates to that $H$ must have at least one eigenstate, i.e., a nonempty point spectrum.

\end{enumerate}

%==========================================================
\vspace{0.0mm}\noindent\textbf{A.2. Purifications in Hilbert-Schmidt bundle}\vspace{0.5mm}\\
\csname phantomsection\endcsname%
\addcontentsline{toc}{subsection}{A.2. Purifications in Hilbert-Schmidt bundle}%
\label{Part:A2}%
\makeatletter\def\zref@titleref@current{A.2}\makeatother
\zlabel{Part:A2}%
%==========================================================
%
We study the Hilbert-Schmidt bundle of purifications $A$ of density operators $\rho$ on our Hilbert space $\cH$.
The goal is to derive Eq.~\eqref{cGcF_XY} for the Bures metric and the Uhlmann curvature.
To avoid technical issues, we give the presentation for the case when $\dim(\cH) < \infty$, along the lines of \cite{BeZy:2009} for the main part.
We then explain in what sense the final formulas hold for the $\dim(\cH) = \infty$ case in Part~\hyperref[Part:A3]{\ztitleref{Part:A3}} below.
Other metric-curvature pairs are briefly discussed in Part~\hyperref[Part:A6]{\ztitleref{Part:A6}}.

To begin, consider full-rank operators $A$ in the Hilbert space $\cH \otimes \cH^\ast$ of Hilbert-Schmidt operators that project under $\pi: A \mapsto \rho = AA^\dagger$ to the space of positive density operators $\rho > 0$.\footnote{Operators are taken to have full rank (trivial kernel) in order for the fibers above different $\rho$ to be isomorphic to each other.}
This is a fiber bundle with the interior of $\cD$ as its base space; the boundary $\partial\cD$ can later be recovered by continuity \cite{BeZy:2009}.
Purifications of $\rho$ are only unique up to unitary operators $U$, since all $A = \sqrt{\rho}U$ satisfy $\rho = AA^\dagger$.
This says that the fiber above each $\rho$ is the space of all $\sqrt{\rho}U$, which thus is isomorphic to the unitary group acting from the right on a given $A$.
Consequently, our bundle is a principal fiber bundle with the unitary operators as its structure group, and since there is a \emph{global section}\footnote{The square root is unique for positive operators.} $s: \rho \mapsto \sqrt{\rho}$, it is even a trivial bundle.
(The last observation will be of no consequence here.)

Tangent vectors $v$ to the bundle at a point $A$ are also Hilbert-Schmidt operators, with any \emph{vertical} vector written as $AW$ for some anti-Hermitian operator $W$.\footnote{Any curve along a fiber through a point $A$ must be of the form $A\ee^{Wt} = A + AWt + \ldots$ for $t \in \mathbb{R}$.}
This reflects that the vertical subspace is isomorphic to the Lie algebra of the structure group.
We will sometimes write tangent vectors $v$ using the 1-form $\dd A$, whose action on $v$ gives $\dd A(v) = v$, and tangent vectors $X$ to the base space using $\dd \rho$.
With that in mind, in reference to the real part of the Hilbert-Schmidt inner product $A \cdot B = \frac{1}{2} \Tr (A^\dagger B + B^\dagger A)$, the $\dd A$ decomposes as\footnote{To prove Eq.~\eqref{ddA_eq}, we first write it using tangent vectors: $v = \dd A(v) = G(v)A + \ii AH(v)$.
The $v$ can be any Hilbert-Schmidt operator, the space of which has real dimension $2N^2$.
The vertical component $v|_V$ of $v$ was shown to be $\ii AH(v)$ for some Hermitian operator $H(v)$, and the space of those has real dimension $N^2$.
This leaves a space of dimension $2N^2 - N^2 = N^2$. 
Now, the horizontal component $v|_H$ must be orthogonal to all $\ii AH(v)$, i.e., $0 = \Tr \ii \bigl( v|_H^\dagger A - A^\dagger v|_H\pdag \bigr) H(v)$ for all $H(v)$, so $v|_H^\dagger A - A^\dagger v|_H\pdag = 0$.
This is solved by $v|_H = G(v)A$ for some Hermitian operator $G(v)$, and since this spans the whole missing space of dimension $N^2$, all $v|_H$ must have that form.}
\begin{equation}
\label{ddA_eq}
\dd A = G A + \ii A H
\end{equation}
using 1-forms $G$, $H$ whose values $G(v)$, $H(v)$ are Hermitian operators.
This defines an Ehresmann connection, in that $\dd A \mapsto GA$ defines a projection onto the \emph{horizontal} tangent bundle, or equivalently, $\dd A \mapsto \ii AH$ defines a projection onto the \emph{vertical} tangent bundle.
The value of each is naturally Hilbert-Schmidt.

In what follows, let $X$, $Y$ be tangent vectors at a given $\rho$, and let $v_X$, $v_Y$ be any tangent vectors at a purification $A$ of $\rho$ that ``project'' to $X$, $Y$, i.e., are \emph{lifts} of $X$, $Y$.
We have $\dd \rho = \dd (AA^\dagger) = (\dd A) A^\dagger + A \dd A^\dagger = G AA^\dagger + AA^\dagger G = G \rho + \rho \, G$, where we used Eq.~\eqref{ddA_eq}.
Written not using 1-forms $\dd\rho$, $\dd A$, but tangent vectors $X$, $v_X$, this shows that the value $G(v_X)$ depends only on $X$, hence the notation $G_{X} \equiv G(v_X)$, and we get precisely Eq.~\eqref{Lyapunov_eq}---this is the origin of that equation.

The Bures metric is obtained from minimizing the squared norm \cite{BeZy:2009}
\begin{equation}
\label{Bures_minimization}
v_X\pdag \cdot v_X\pdag = \Tr v_X^\dagger v_X\pdag
\end{equation}
over all lifts $v_X$ of $X$.
Using $v_X = \dd A(v_X)$ and the decomposition \eqref{ddA_eq}, it follows that the minimum occurs at the \emph{horizontal lift}, i.e., when $\dd A = G A$.
Using the usual bijection between a symmetric bilinear form $B(u,v) = \frac{1}{2} [Q(u+v) - Q(u) - Q(v)]$ and its quadratic form $Q(u) = B(u,u)$, the corresponding metric is given by
\begin{equation}
\frac{1}{2} \Tr \bigl[ v_X^\dagger v_Y\pdag + v_Y^\dagger v_X\pdag \bigr]
\end{equation}
for $v_X = G_X A$ and $v_Y = G_Y A$.
This simplifies to $\frac{1}{2} \Tr \rho \, \{ G_X, G_Y \}$, which is a real value, independent of the choice of purification.
It can thus be assigned to the base space, yielding Eq.~\eqref{cG_XY}.

The (scalar) Uhlmann curvature is obtained from the curvature \cite{KoMiSl:1993}
\begin{equation}
R(v_X, v_Y) = \bigl[ v_X|_{\mathrm{H}}, v_Y|_{\mathrm{H}} \bigr] \big|_{\mathrm{V}}
\end{equation}
of the bundle, where $|_{\mathrm{H}}$ and $|_{\mathrm{V}}$ indicate projections onto the horizontal and vertical tangent bundle, respectively (see above).
We will view it as a functional on $\cH \otimes \cH^\ast$ with respect to the Hilbert-Schmidt inner product, $\Tr R(v_X, v_Y)^\dagger A \in \mathbb{C}$.
From Eq.~\eqref{ddA_eq}, we can identify the Uhlmann connection 1-form $\omega \equiv \ii H$, whose value acts from the right on $A$.
Then, in the conventions of \cite{KoMiSl:1993}, $R = - A\Omega$ using the curvature form $\Omega = \dd\omega + \omega \wedge \omega = \ii (\dd H + \ii H \wedge H)$, where $\dd$ denotes exterior derivative and $\wedge$ is the usual wedge product.\footnote{Herein lies an issue for the $\dim(\cH) = \infty$ case. These equalities and manipulations would only be formal since the operators $H$ would not necessarily be bounded, and thus neither would $\Omega$, nor then would it be Lie algebra valued. In the end, however, all we need is $A\Omega A^\dagger$, which turns out to be trace class and thus bounded.}
All we need is the combination $A\Omega A^\dagger$, which is fortunate since $H$ is only known in terms of the combination $\ii AH$. 
To compute $A\Omega A^\dagger$, we follow \cite{Uhlmann:1991} and start from the decomposition \eqref{ddA_eq} and its conjugate version $\dd A^\dagger = A^\dagger G - \ii H A^\dagger$.
One finds
\begin{align}
0
& = \dd^2 A = (\dd G)A - G \wedge \dd A + \ii \dd A \wedge H + \ii A \dd H \nonumber \\
& = (\dd G - G \wedge G) A + A \Omega,
\end{align}
and similarly
\begin{equation}
0 = \dd^2 A^\dagger = A^\dagger (\dd G + G \wedge G) - \Omega A^\dagger,
\end{equation}
where we used $\dd^2 = 0$, the graded product rule, and the decompositions of $\dd A$ and $\dd A^\dagger$ once more.
Multiplying the first equation by $A^\dagger$ from the right and the second by $A$ from the left, subtracting one from the other, and taking the trace, we obtain $\Tr A \Omega A^\dagger = \Tr AA^\dagger(G\wedge G)$.
Now, using that $\Omega^\dagger = - \Omega$, we have $\Tr R^\dagger A = - \Tr \Omega^\dagger A^\dagger A = \Tr A \Omega A^\dagger = \Tr \rho \, (G \wedge G)$.
Consequently, $\Tr R(v_X, v_Y)^\dagger A = \Tr \rho \, [G_X, G_Y]$, which is a purely imaginary value, independent of the choice of purification.
As for the metric, the curvature can thus also be assigned to the base space, giving Eq.~\eqref{cF_XY} after multiplication by $-\ii$ to get a real value.

Note that we never used the existence of a global section.
While its presence or absence is important for nontrivial topological invariants, cf.\ \cite{BudichDiehl:2015}, what mattered here was to define geometric quantities on the base space from geometric quantities on the purification bundle that are \emph{gauge invariant}, i.e., independent of the choice of purification, so local sections suffice.

%==========================================================
\vspace{0.0mm}\noindent\textbf{A.3. Tangent vectors when $\boldsymbol{\dim(\cH) = \infty}$}\vspace{0.5mm}\\
\csname phantomsection\endcsname%
\addcontentsline{toc}{subsection}{A.3. Tangent vectors when \texorpdfstring{$\dim(\cH) = \infty$}{the dimension of \emph{H} is infinite}}%
\label{Part:A3}%
\makeatletter\def\zref@titleref@current{A.3}\makeatother
\zlabel{Part:A3}%
%==========================================================
%
If $\cH$ is a separable infinite-dimensional Hilbert space, one must further restrict the allowed tangent vectors $X$ at a point $\rho \in \cD$.
The requirements are that $X$ can be any traceless Hermitian operator\footnote{$X$ is trace class, thus bounded, and so it is self-adjoint and not only Hermitian.} on $\cH$ so that:
(i) $P_{\hspace{1pt}\ker \rho} X P_{\hspace{1pt}\ker \rho} = 0$, where $P_{\hspace{1pt}\ker \rho}$ is the projector onto $\ker \rho$.
(ii) $G_X \sqrt{\rho}$ is a Hilbert-Schmidt operator, where $G_X$ is the solution to Eq.~\eqref{Lyapunov_eq} interpreted as $X = (G_X\sqrt{\rho})\sqrt{\rho} + \sqrt{\rho}(G_X \sqrt{\rho})^\dagger$ with $G_X^\dagger = G_X$ on $\ran \sqrt{\rho} \equiv \{ \hspace{0.5pt} \sqrt{\rho} |\psi\rangle \in \cH : |\psi\rangle \in \cH \hspace{0.5pt} \}$.
Specifically, the Hilbert-Schmidt condition says that $\Tr |G_X \sqrt{\rho}|^2 = \Tr (G_X \sqrt{\rho})^\dagger G_X \sqrt{\rho} = \Tr \rho G_{X}^2 < \infty$, and it comes from the construction on the Hilbert-Schmidt bundle [cf.\ Eq.~\eqref{ddA_eq}] since $GA$ must be Hilbert-Schmidt and $A = \sqrt{\rho} U$.
This ensures that $\cG_{\rho}(X,Y)$ and $\cF_{\rho}(X,Y)$ in Eq.~\eqref{cGcF_XY} are well defined (each term is a trace of a product of Hilbert-Schmidt operators).
Moreover, one can show that $\Tr \rho G_{K_\rho X}^2 \leq \Tr \rho G_{X}^2$ [Eq.~\eqref{GX_X_ineq} in Part~\hyperref[Part:A4]{\ztitleref{Part:A4}}], so $G_{K_\rho X} \sqrt{\rho}$ is then also Hilbert-Schmidt.

The observables $\cO_a$ introduced when discussing linear responses and fluctuations in the main text are also required to belong to certain families, depending on the dynamics.
Recall that any dynamical map $\Phi_t$ is required to have at least one stationary $\rho \in \cD$, i.e., $\Phi_t \rho = \rho$.
Let $\rho$ be a given such state.

\noindent
{\bf Definition 1.}
\csname phantomsection\endcsname%
\label{Def:1}%
\makeatletter\def\zref@titleref@current{1}\makeatother
\zlabel{Def:1}%
Let $\cA_{\rho}$ be the family of all (unbounded) self-adjoint operators $\cO$ on $\cH$ such that  $\cO \sqrt{\rho}$ is Hilbert-Schmidt.

\noindent
{\bf Definition 2.}
\csname phantomsection\endcsname%
\label{Def:2}%
\makeatletter\def\zref@titleref@current{2}\makeatother
\zlabel{Def:2}%
Let $\cB$ be the family of all bounded self-adjoint operators $\cO$ on $\cH$.\footnote{Note that $\cO$ bounded means that $\cO \sqrt{\rho}$ is Hilbert-Schmidt since $\sqrt{\rho}$ is Hilbert-Schmidt.}

For both families, we impose zero averages, $\Tr \rho \, \cO = 0$, which can always be satisfied by replacing $\cO$ by $\cO - \Tr \rho \, \cO $.
The key is that such a recentering only takes place for the initial observable at $t = 0$, and the dynamics needs to preserve the zero average.

Definitions~\hyperref[Def:1]{\ztitleref{Def:1}} and~\hyperref[Def:2]{\ztitleref{Def:2}} are motivated by that each $\XO_a$ defined as in Eq.~\eqref{Xa_t} for a certain $\cO_a$ must be an allowed tangent vector.
It remains to discuss in what sense the dynamics ensures that $\XO_a(t)$ is an allowed tangent vector for all $t \geq 0$.

First, we consider unitary dynamics, for which one can straightforwardly prove the following:

\noindent
{\bf Lemma 1.}
\csname phantomsection\endcsname%
\label{Lemma:1}%
\makeatletter\def\zref@titleref@current{1}\makeatother
\zlabel{Lemma:1}%
For $\Phi_t = \ee^{-\ii t [H, \bullet]}$ generated by a Hamiltonian $H$ with a stationary $\rho \in \cD$, given a $\cO \in \cA_{\rho}$ (or $\cB$) with $\Tr \rho \, \cO = 0$, then $\cO (t) \equiv \ee^{\ii Ht} \cO \ee^{-\ii Ht} \in \cA_{\rho}$ (or $\cB$) with $\Tr \rho \, \cO(t) = 0$ for all $t \geq 0$.

Second, we turn to trace-preserving completely positive (TPCP) maps $\Phi_t$, such as the Markovian case of Lindblad dynamics \cite{Lindblad:1976}.
To proceed, we need to introduce the adjoint map $\Phi_t^*$ defined by $\Tr (\Phi_t \sigma) \cO = \Tr \sigma \Phi_t^* \cO$ for all trace-class operators $\sigma$ and bounded operators $\cO$.
This allows us to define $\cO(t) \equiv \Phi_t^* \cO$, which are thus also bounded, which is why one typically needs to consider the family $\cB$ (see Lemma~\hyperref[Lemma:2]{\ztitleref{Lemma:2}} below).
To circumvent this requirement needs refinements, which we also give (see Lemma~\hyperref[Lemma:3]{\ztitleref{Lemma:3}} below).

To begin, since $\Phi_t$ preserves Hermiticity, so does $\Phi_t^*$.
Thus, $\cO(t)^\dagger = (\Phi_t^* \cO)^\dagger = \Phi_t^* \cO^\dagger = \Phi_t^* \cO = \cO(t)$.
It remains to verify that zero average is preserved, which it is: $\Tr \rho \, \cO(t) = \Tr \rho \, \Phi_t^* \cO = \Tr \, (\Phi_t \rho) \, \cO = \Tr \rho \, \cO = 0$.
We have thus proved the following:

\noindent
{\bf Lemma 2.}
\csname phantomsection\endcsname%
\label{Lemma:2}%
\makeatletter\def\zref@titleref@current{2}\makeatother
\zlabel{Lemma:2}%
For a TPCP map $\Phi_t$ with a stationary $\rho \in \cD$, given $\cO \in \cB$ with $\Tr \rho \, \cO = 0$, then $\cO(t) \in \cB$ with $\Tr \rho \, \cO(t) = 0$ for all $t \geq 0$.

It is possible to relax to unbounded self-adjoint operators $\cO$ at time $t = 0$ by defining $\cO(t)$ for $t > 0$ from $\Phi_t^*$ acting on $\cO_\Lambda \equiv \cO P_{[-\Lambda, \Lambda]} - \Tr \rho \, \cO P_{[-\Lambda, \Lambda]}$, where $P_{[-\Lambda, \Lambda]}$ is the spectral projection onto $[-\Lambda, \Lambda]$ for some sharp cutoff $\Lambda > 0$.
We cannot guarantee that such $\cO(t)$ will be self-adjoint, and we instead show that $\cO(t)^\dagger = \cO(t)$ on $\ran \sqrt{\rho}$ in a Hilbert-Schmidt sense.
However, that is what we need for our applications in Eqs.~\eqref{kappa_ab_result} and \eqref{fluct_cG}, with the right-hand sides interpreted as $2 \Im \Tr \, [\cO_a(t)\sqrt{\rho}]^\dagger \cO_b \sqrt{\rho}$ and $\Re \Tr \, [\cO_a(t)\sqrt{\rho}]^\dagger \cO_b \sqrt{\rho}$, respectively, and Eq.~\eqref{Xa_t} interpreted as $\XO_a(t) \equiv [\cO_a(t) \sqrt{\rho}]\sqrt{\rho} + \sqrt{\rho} [\cO_a(t)\sqrt{\rho}]^\dagger$.
Indeed, since $\Phi_t^*$ is positive and unital,\footnote{I.e., $\Phi_t^* \mathrm{Id} = \mathrm{Id}$, which comes from that $\Phi_t$ is trace-preserving.} by the Kadison-Schwarz inequality \cite{Kadison:1952}, $\Tr \, ([\Phi_t^* B] \sqrt{\rho})^\dagger [\Phi_t^* B]\sqrt{\rho} = \Tr \rho \, (\Phi_t^* B)^2 \leq \Tr \rho \, \Phi_t^* B^2 = \Tr \, (\Phi_t \rho) B^2 = \Tr \rho \, B^2 = \Tr \, (B \sqrt{\rho})^\dagger B\sqrt{\rho}$ for any bounded self-adjoint operator $B$.
Applied to $B = \cO_{\Lambda_m} - \cO_{\Lambda_n}$ for any Cauchy sequence $\cO_{\Lambda_n} \sqrt{\rho}$ ($n = 1, 2, \ldots$) in Hilbert-Schmidt norm, it follows that $[\Phi_t^* \cO_{\Lambda_n}] \sqrt{\rho}$ is also Cauchy.
Using completeness of Hilbert-Schmidt space, we infer that $\cO(t) \sqrt{\rho}$ defined as the limit of $[\Phi_t^* \cO_{\Lambda_n}] \sqrt{\rho}$ as $\Lambda_n \to \infty$ exists in said space for all $t \geq 0$.
Combined with our previous computations, we have proved the following:

\noindent
{\bf Lemma 3.}
\csname phantomsection\endcsname%
\label{Lemma:3}%
\makeatletter\def\zref@titleref@current{3}\makeatother
\zlabel{Lemma:3}%
For a TPCP map $\Phi_t$ with a stationary $\rho \in \cD$, given $\cO \in \cA_\rho$ with $\Tr \rho \, \cO = 0$, define $\cO(t)$ through its action $\cO(t) \sqrt{\rho} \equiv \lim_{\Lambda \to \infty} [\Phi_t^* \cO_{\Lambda}] \sqrt{\rho}$.
Then $\cO(t)^\dagger = \cO(t)$ on $\ran \sqrt{\rho}$, $\Tr \rho \, \cO(t) \equiv \Tr \sqrt{\rho} \cO(t) \sqrt{\rho} = 0$ and $\cO(t) \sqrt{\rho}$ is Hilbert-Schmidt for all $t \geq 0$.

%==========================================================
\vspace{0.0mm}\noindent\textbf{A.4. Mixed-state geometry using matrices}\vspace{0.5mm}\\
\csname phantomsection\endcsname%
\addcontentsline{toc}{subsection}{A.4. Mixed-state geometry using matrices}%
\label{Part:A4}%
\makeatletter\def\zref@titleref@current{A.4}\makeatother
\zlabel{Part:A4}%
%==========================================================
%
Let us see what the metric, the curvature, and the fluctuation-dissipation structure in Eqs.~\eqref{cGcF_XY}--\eqref{Riesz_rep_eq} are for a quantum system with dimension $\dim(\cH) \equiv N$ when expressed using a basis.
The presentation covers both $N < \infty$ and $N = \infty$ (with the conditions in Part~\hyperref[Part:A3]{\ztitleref{Part:A3}}).
One can also write everything explicitly without a basis by using, e.g., semigroups \cite{Pazy:1983}, but we will not present that here.

Since $\rho \in \cD$, there is a (nonunique) basis $|e_j\rangle$ so that $\rho = \sum_{j} \prho_j |e_j\rangle \langle e_j|$ is diagonal,
with $\prho_j \geq 0$ satisfying $\sum_j \prho_j = 1$.\footnote{Depending on whether $N < \infty$ or $N = \infty$, the indices run over $\{ 1, \ldots, N \}$ or $\mathbb{N} = \{ 1, 2, \ldots \}$ and $\sum_j \equiv \sum_{j=1}^{N}$ or $\sum_j \equiv \sum_{j=1}^{\infty}$, unless specified otherwise.}
In this basis, the tangent vectors are $X = \sum_{j,k} X_{jk} |e_j\rangle \langle e_k|$ with $X_{jk} = \overline{X_{kj}}$ satisfying $\sum_j X_{jj} = 0$
and $\Tr |G_X \sqrt{\rho}|^2 = \sum_{j,k} \prho_j |(G_X)_{jk}|^2 < \infty$,
where $G_{X}$ solves Eq.~\eqref{Lyapunov_eq}.
The latter is given by
\begin{equation}
\label{G_jk}
\bigl( G_{X} \bigr)_{jk} = \frac{1}{\prho_j + \prho_k} X_{jk},
\end{equation}
which inserted into Eq.~\eqref{cGcF_XY} yields
\begin{subequations}
\label{cGcF_XY_jk}
\begin{align}
\cG_{\rho}(X,Y)
& = \frac{1}{2} \sum_{j,k} \frac{1}{\prho_j + \prho_k} \overline{X_{jk}} Y_{jk}, \label{cG_XY_jk} \\
\cF_{\rho}(X,Y)
& = \ii \sum_{j,k} \frac{\prho_j - \prho_k}{(\prho_j + \prho_k)^2} \overline{X_{jk}} Y_{jk}. \label{cF_XY_jk}
\end{align}
\end{subequations}
Moreover, from Eq.~\eqref{Riesz_rep_eq}, we find
\begin{equation}
\label{Krho_jk}
\bigl( K_{\rho} X \bigr)_{jk} = - \ii \frac{\prho_{j} - \prho_{k}}{\prho_{j} + \prho_{k}} X_{jk},
\end{equation}
giving the action of $K_{\rho}$ on any allowed $X$.
This could also have been obtained directly from Eqs.~\eqref{FDS} and \eqref{cGcF_XY_jk} using that the symmetric form $\cG_\rho$ in Eq.~\eqref{cG_XY_jk} is nondegenerate, with Eq.~\eqref{Riesz_rep_eq} then obtained as the basis-independent formulation of Eq.~\eqref{Krho_jk}.
Meanwhile, the numerators in Eq.~\eqref{cF_XY_jk}--\eqref{Krho_jk} make clear that the antisymmetric form $\cF_{\rho}$ indeed can be degenerate and that $K_{\rho}$ is generally not invertible.

It is straightforward to check that $K_{\rho} X$ in Eq.~\eqref{Krho_jk} is a tangent vector.
Indeed, $\Tr K_{\rho} X = \sum_{j} (K_{\rho} X)_{jj} = 0$ and $\overline{(K_{\rho} X)_{jk}} = (K_{\rho} X)_{kj}$, along with
\begin{align}
\label{GX_X_ineq}
\Tr \rho \, G_{K_\rho X}^2
& = \Tr \rho G_{X}^2 - \sum_{j,k} \frac{2\prho_j\prho_k}{(\prho_j + \prho_k)^3} |X_{jk}|^2 \nonumber \\
& \leq \Tr \rho \, G_{X}^2 < \infty.
\end{align}
Note that we have equality if and only if all terms in $\sum_{j,k} \frac{2\prho_j\prho_k}{(\prho_j + \prho_k)^3} |X_{jk}|^2$ are zero.
Since $X_{jk}$ is only allowed to be nonzero if one of $\prho_j$ and $\prho_k$ is nonzero, the only way all terms are zero is if exactly one eigenvalue of $\rho$ is nonzero.
I.e., $\Tr \rho \, G_{K_\rho X}^2 = \Tr \rho \, G_{X}^2$ for all allowed $X$ if and only if $\rho$ is pure.

To get an understanding of how the formulas above work, we consider two special cases:

\noindent
{\bf Pure states.}
Let $\rho_{\psi} \equiv |\psi\rangle\langle\psi|$ with $\langle\psi|\psi\rangle = 1$.
Without loss of generality, take $\prho_1 = 1$ and $\prho_{j\neq1} = 0$.
Since any tangent vector $X$ at this $\rho_{\psi}$ can only point from the $|e_1\rangle$ direction to one of the others, we must have $X_{jk} = 0$ unless exactly one of $j,k$ is $1$ ($X_{11} = 0$ since $\Tr X = 0$).
Eqs.~\eqref{cGcF_XY_jk}--\eqref{Krho_jk} then become
\begin{subequations}
\label{cGcF_XY_pure_jk}
\begin{align}
\cG_{\rho_{\psi}}(X,Y)
& = \frac{1}{2} \sum_{k \geq 2} \bigl(\hspace{1pt} \overline{X_{1k}} Y_{1k} + \overline{X_{k1}} Y_{k1} \bigr), \label{cG_XY_pure_jk} \\
\cF_{\rho_{\psi}}(X,Y)
& = \ii \sum_{k \geq 2} \bigl(\hspace{1pt} \overline{X_{1k}} Y_{1k} - \overline{X_{k1}} Y_{k1} \bigr), \label{cF_XY_pure_jk}
\end{align}
\end{subequations}
and the only nonzero elements of $K_{\rho_{\psi}} X$ are
\begin{equation}
\label{Krho_pure_jk}
\bigl( K_{\rho_{\psi}} X \bigr)_{k1} = \ii X_{k1},
\qquad
\bigl( K_{\rho_{\psi}} X \bigr)_{1k} = -\ii X_{1k},
\end{equation}
so $K_{\rho_{\psi}}^2 X = -X$.
The $\cG_{\rho_{\psi}}$ and $\cF_{\rho_{\psi}}$ above  are the Fubini-Study metric and symplectic form,\footnote{In bra-ket notation, one has $X = |\dd\psi_X\rangle\langle\psi| + |\psi\rangle\langle\dd\psi_X|$ with $|\dd\psi_X\rangle = \sum_{k} \overline{X_{1k}} |e_{k}\rangle$, and similarly for $Y$. 
Eq.~\eqref{cG_XY_pure_jk} then becomes
$\cG_{\rho_{\psi}}(X,Y) = \bigl( \langle\dd\psi_X|\dd\psi_Y\rangle + \langle\dd\psi_Y|\dd\psi_X\rangle \bigr)/2 - \langle\dd\psi_X|\psi\rangle \langle\psi|\dd\psi_Y\rangle$ and Eq.~\eqref{cF_XY_pure_jk} becomes
$\cF_{\rho_{\psi}}(X,Y) = -\ii \bigl( \langle\dd\psi_X|\dd\psi_Y\rangle - \langle\dd\psi_Y|\dd\psi_X\rangle \bigr)$.}
and the map $K_{\rho_{\psi}}$ is the complex structure $J$ for pure states.
Thus, $2\cG_{\rho_{\psi}}$, $\cF_{\rho_{\psi}}$, and $K_{\rho_{\psi}} = J$ form a compatible triplet, which together with integrability of $J$ is the statement that the manifold of pure states is K\"ahler.

Comparing Eqs.~\eqref{cGcF_XY} and \eqref{cGcF_XY_pure_jk}, one sees that $G_{X} = X$ contains all relevant information from Eq.~\eqref{Lyapunov_eq}.
If one drops the criterion $X_{11} = 0$ by replacing $X$ by $X - \langle e_1|X|e_1\rangle$, then the metric can be written as
\begin{equation}
\label{PV_eq}
\cG_{\rho_{\psi}}(X,Y)
= \Re \langle\psi|XY|\psi\rangle - \langle\psi|X|\psi\rangle \langle\psi|Y|\psi\rangle.
\end{equation}
This is the pure-state version of the fluctuation interpretation \eqref{fluct_cG}, as it is the real part of the covariance $\operatorname{Cov}_{\rho_{\psi}}(X, Y) = \langle\psi|XY|\psi\rangle - \langle\psi|X|\psi\rangle \langle\psi|Y|\psi\rangle$.
On parameter spaces, such a formula for the quantum metric and its fluctuation interpretation appeared much earlier in \cite{ProvostVallee:1980}.

\noindent
{\bf Thermal states.}
Let $\rho_{\beta} \equiv Z_{\beta}^{-1} \sum_{j} \ee^{-\beta E_j} |e_j\rangle\langle e_j|$ at inverse temperature $\beta > 0$ with partition function $Z_{\beta} \equiv \sum_{j} \ee^{-\beta E_j} < \infty$.
Here, $E_j$ are the system's eigenenergies.
Clearly, $\prho_j = Z_{\beta}^{-1} \ee^{-\beta E_j}$, which inserted into Eq.~\eqref{cGcF_XY_jk} yields the metric $\cG_{\rho_{\beta}}$ and curvature $\cF_{\rho_{\beta}}$.
More revealing is what Eq.~\eqref{Krho_jk} becomes:
\begin{equation}
\label{Krho_beta_jk}
\bigl( K_{\rho_{\beta}} X \bigr)_{jk} = \ii \tanh \bigl( \beta \omega_{jk}/2 \bigr) X_{jk}
\end{equation}
with $\omega_{jk} \equiv E_j - E_k$.
Inserted into Eq.~\eqref{FDS}, it gives the distribution function $\frac{1}{2} \coth \bigl( \beta \omega_{jk}/2 \bigr)$ that appears in the usual fluctuation-dissipation relation \cite{CallenWelton:1951, Kubo:1957, Kubo:1966}.

\noindent
{\bf Fluctuations.}
By splitting Eq.~\eqref{cG_XY_jk} into its (classical) Fisher-Rao and quantum terms \cite{BeZy:2009},
\begin{equation}
\label{cG_XY_jk_split}
\cG_{\rho}(X, Y) = \frac{1}{4} \sum_{j} \frac{X_{jj}Y_{jj}}{\prho_j} + \frac{1}{2} \sum_{j \neq k} \frac{\overline{X_{jk}} Y_{jk}}{\prho_j + \prho_k},
\end{equation}
one sees that there are only quantum fluctuations at pure $\rho$.
Classical statistical fluctuations take over the closer $\rho$ gets to the maximally mixed state $\rho_{\ast}$, at which point quantum fluctuations vanish (seen by making a change of basis so that $X$ is diagonal).
Note that $\rho_{\ast} \equiv \frac{1}{N} \mathrm{Id}$ lies in $\cD$ for $N < \infty$, but not for $N = \infty$.
As stated in the main text, since $(K_{\rho} X)_{jj} = 0$ due to Eq.~\eqref{Krho_jk}, only the quantum part of the metric matters in Eq.~\eqref{FDS}, verifying that the curvature is quantum.

%==========================================================
\vspace{0.0mm}\noindent\textbf{A.5. Proof of Theorems 1 and 2}\vspace{0.5mm}\\
\csname phantomsection\endcsname%
\addcontentsline{toc}{subsection}{A.5. Proof of Theorems 1 and 2}%
\label{Part:A5}%
\makeatletter\def\zref@titleref@current{A.5}\makeatother
\zlabel{Part:A5}%
%==========================================================
%
As stated before, Eqs.~\eqref{G_jk}--\eqref{Krho_jk} hold for separable Hilbert spaces (cf.\ Part~\hyperref[Part:A3]{\ztitleref{Part:A3}}) and not only finite-dimensional ones.
They thus allow one to prove Theorems~\hyperref[Theorem:1]{\ztitleref{Theorem:1}} and~\hyperref[Theorem:2]{\ztitleref{Theorem:2}}.
For Theorem~\hyperref[Theorem:1]{\ztitleref{Theorem:1}}, this was already done below Eq.~\eqref{Krho_jk}, while here we give a different proof that does not use decompositions with respect to a basis.
Alternatively, one can also prove them using semigroup techniques.

\begin{proof}[Proof of Theorem 1]
From Eqs.~\eqref{cGcF_XY} and \eqref{FDS}, we obtain
\begin{equation}
\label{thm1_proof}
\Tr \Bigl[ \rho \bigl( G_{K_\rho X} + \ii G_X \bigr) + \bigl( G_{K_\rho X} - \ii G_X \bigr) \rho \Bigr] G_{Y} = 0,
\end{equation}
for all $X$, $Y$.
Let $\tilde{X}$ denote the expression within the square brackets above.
Clearly, $\tilde{X}^\dagger = \tilde{X}$ and $\Tr \tilde{X} = 2 \Tr \rho \, G_{K_\rho X} = \Tr K_\rho X = 0$ using Eq.~\eqref{Lyapunov_eq}.
Introduce $G_{\tilde{X}}$ as the solution to Eq.~\eqref{Lyapunov_eq} for $\tilde{X}$.
Eq.~\eqref{thm1_proof} then becomes $\Tr \rho \, \{ G_{\tilde{X}}, G_{Y} \} = 0$.
(Setting $Y = \tilde{X}$, this says that $G_{\tilde{X}} \sqrt{\rho}$ is Hilbert-Schmidt, verifying that $\tilde{X}$ is a tangent vector to $\cD$ at $\rho$.)
In other words, Eq.~\eqref{thm1_proof} says that the Bures metric $\cG_{\rho}(\tilde{X},Y) = 0$, and since $\cG_{\rho}$ is nondegenerate, we must have $\tilde{X} = 0$.
Eq.~\eqref{Lyapunov_eq} then implies that
$K_\rho X = \rho G_{K_\rho X} + G_{K_\rho X} \rho = - \ii \rho G_X + \ii G_X \rho$,
from which Eq.~\eqref{Riesz_rep_eq} follows by inserting the above and using Eq.~\eqref{Lyapunov_eq} one more time.
\end{proof}

\begin{proof}[Proof of Theorem 2]
We prove that, under the same conditions, $K_{\rho}$ is almost complex.\footnote{The manifold of pure states, the projective Hilbert space $\mathbb{P} \cH$, is well known to be K\"ahler, thus its (almost) complex structure is integrable.} I.e., $K_{\rho}^2 X = -X$ if and only if $\rho$ is pure.
This holds if and only if all factors in Eq.~\eqref{Krho_jk} multiplying a nonzero $X_{jk}$ satisfy
\begin{equation}
\biggl( \frac{\prho_{j} - \prho_{k}}{\prho_{j} + \prho_{k}} \biggr)^2 = 1.
\end{equation}
It is clearly true when only one of the eigenvalues is nonzero, i.e., $\rho$ is pure.
Thus, assume that at least two are nonzero, say, $\prho_1$ and $\prho_2$.
Then either $\prho_1 + \prho_2 = \prho_1 - \prho_2$ or $\prho_1 + \prho_2 = - \prho_1 + \prho_2$, meaning that $\prho_2 = 0$ or $\prho_1 = 0$, contradicting the assumption.
\end{proof}

%==========================================================
\vspace{0.0mm}\noindent\textbf{A.6. Other metric-curvature pairs}\vspace{0.5mm}\\
\csname phantomsection\endcsname%
\addcontentsline{toc}{subsection}{A.6. Other metric-curvature pairs}%
\label{Part:A6}%
\makeatletter\def\zref@titleref@current{A.6}\makeatother
\zlabel{Part:A6}%
%==========================================================
%
All that we used to define the structure \eqref{FDS} was nondegeneracy of the metric.
As such, our approach can be applied to explore the physical meaning of similar structures connecting any Riemannian metric on the space of density operators and any (scalar) curvature originating from some connection on the bundle of quantum state purifications.
While the Bures metric is special to us, there is an infinite number of monotone Riemannian metrics on $\cD$, all classified in terms of operator monotone functions if $\cH$ is finite dimensional  \cite{MorozovaChentsov:1989, Petz:1996}, including well-known ones like the Kubo-Mori metric or the Wigner-Yanase metric \cite{GibiliscoIsola:2003}.
Connections corresponding to other metrics have been worked out in \cite{DittmannRudolph:1992, DittmannUhlmann:1999}, but we are not aware of any explicit result for another curvature of the purification bundle than the Uhlmann curvature.

\vspace{2mm}

%==========================================================
\vspace{0.0mm}\noindent\textbf{\fontsize{11}{13.5}\selectfont B.\, Linear response theory}\vspace{0.5mm}
\csname phantomsection\endcsname%
\addcontentsline{toc}{section}{B. Linear response theory}%
\label{Part:B}
\makeatletter\def\zref@titleref@current{B}\makeatother
\zlabel{Part:B}%
%==========================================================

\vspace{-1.5mm}

%==========================================================
\vspace{0.0mm}\noindent\textbf{B.1. Linear response for closed systems}\vspace{0.5mm}\\
\csname phantomsection\endcsname%
\addcontentsline{toc}{subsection}{B.1. Linear response for closed systems}%
\label{Part:B1}%
\makeatletter\def\zref@titleref@current{B.1}\makeatother
\zlabel{Part:B1}%
%==========================================================
%
Consider a Hamiltonian $H$ for a system in a stationary initial state $\rho \in \cD$ at times $t < t_0$ for some $t_0$.\footnote{Recall that $H$ must have at least one eigenstate. Moreover, one can set $t_0 = -\infty$ without loss of generality.}
Define the time-dependent Hamiltonian
\begin{equation}
\label{H_lambda}
H(s)
\equiv H - \sum_b \lambda_b(s) \cO_b
\end{equation}
for $\boldsymbol{\lambda}(s) = ( \lambda_b(s) )$ with real-valued functions $\lambda_b(s)$ of time $s$, and consider the time-evolved state
\begin{equation}
\rho(t)
\equiv
U(t) \rho\hspace{1pt} U(t)^{-1},
\quad
U(t)
\equiv
\overset{\longleftarrow}{\cT} \exp \biggl[ - \ii \int_{t_0}^{t} \! \dd s\, H(s) \biggr] \label{U_t_lambda}
\end{equation}
with $\overset{\longleftarrow}{\cT}$ ordering time increasingly from right to left.
We are interested in linear response functions \cite{Kubo:1966}\footnote{The overall sign is as in \cite{Moosavi:iCFT:2024}, and opposite to the sign in \cite{Kubo:1966}.
It reflects that a positive flow is in the direction opposite any gradient in the perturbations.}
\begin{equation}
\label{kappa_ab_def}
\kappa_{ab}(t,s)
\equiv
- \frac{\delta}{\delta \lambda_b(s)} \Tr \bigl[ \rho(t) \cO_a \bigr] \Big|_{\boldsymbol{\lambda}(\cdot) = 0}
\end{equation}
for $s > t_0$.
One obtains $\kappa_{ab}(t,s) = 0$ for $t < s$, and $\kappa_{ab}(t,s) = -\ii \Tr \rho \, \bigl[ \cO_a(t-s), \cO_b \bigr]$ for $t > s$ with time-evolved observables $\cO_a(t) \equiv \ee^{\ii Ht} \cO_a \ee^{-\ii Ht}$.
The result depends only on $t-s$ (since $H$ is time translation invariant).
Defining $\kappa_{ab}(t) \equiv \kappa_{ab}(t,0)$, we obtain
\begin{equation}
\label{kappa_ab_phi}
\kappa_{ab}(t) = \theta(t) \phi_{ab}(t),
\quad
\phi_{ab}(t) \equiv -\ii \Tr \rho \bigl[ \cO_a(t), \cO_b \bigr]
\end{equation}
with the Heaviside function $\theta(t)$.\footnote{$\theta(t) = 1$ for $t > 0$ and $= 0$ for $t < 0$.} This yields
Eq.~\eqref{kappa_ab_result}.
Note that Eqs.~\eqref{kappa_ab_result} and \eqref{kappa_ab_phi} are well-defined for both finite- and infinite-dimensional Hilbert spaces (with the conditions on the observables $\cO_a$ in Part~\hyperref[Part:A3]{\ztitleref{Part:A3}}).
Responses for concrete physical observables are discussed in Part~\hyperref[Part:B3]{\ztitleref{Part:B3}}, and their classification as conductivities or susceptibilities is reviewed in Part~\hyperref[Part:B4]{\ztitleref{Part:B4}}.
Extension beyond linear response is briefly discussed in Part~\hyperref[Part:B7]{\ztitleref{Part:B7}}.

%==========================================================
\vspace{0.0mm}\noindent\textbf{B.2. Linear response for open systems}\vspace{0.5mm}\\
\csname phantomsection\endcsname%
\addcontentsline{toc}{subsection}{B.2. Linear response for open systems}%
\label{Part:B2}%
\makeatletter\def\zref@titleref@current{B.2}\makeatother
\zlabel{Part:B2}%
%==========================================================
%
The derivation above does not require unitary dynamics, just that the perturbations are Hamiltonian.
More generally, we can consider open dynamics under any trace-preserving completely positive map $\Phi_t$, provided it has a stationary $\rho \in \cD$.
For concreteness, we take the Markovian case of Lindblad dynamics \cite{Lindblad:1976}.
That is, $\Phi_t = \ee^{t \cL}$ with the Liouvillian $\cL$ given by
\begin{equation}
\cL\rho = - \ii [H, \rho] + \sum_{j} \Bigl( V_j\pdag \rho V_j^\dagger - \frac{1}{2}\{ V_j^\dagger V_j\pdag, \rho \} \Bigr),
\end{equation}
where $V_j$ are the so-called jump operators.
As before, suppose the system is in a stationary initial state $\rho \in \cD$, i.e., $\Phi_t \rho = \rho$, at times $t < t_0$ for some $t_0$.
We perturb the Hamiltonian as in Eq.~\eqref{H_lambda}, with $\cL(s)$ denoting the corresponding Liouvillian, and consider the time-evolved state
\begin{equation}
\rho(t)
\equiv
\cV(t) \rho,
\quad
\cV(t)
\equiv
\overset{\longleftarrow}{\cT} \exp \biggl[ \int_{t_0}^{t} \! \dd s\, \cL(s) \biggr]
\end{equation}
with the ordering $\overset{\longleftarrow}{\cT}$ as in Eq.~\eqref{U_t_lambda} \cite{BreuerPetruccione:2007}.
Define the linear response functions $\kappa_{ab}(t,s)$ just as in Eq.~\eqref{kappa_ab_def}.
Then, $\kappa_{ab}(t,s) = 0$ for $t < s$, and $\kappa_{ab}(t,s) = - \ii \Tr \Bigl( \ee^{(t-s)\cL} [\cO_b, \bullet] \, \ee^{(s-t_0)\cL} \rho \Bigr) \cO_a = -\ii \Tr \rho \, \bigl[ \cO_a(t-s), \cO_b \bigr]$ for $t > s$, where we used stationarity of $\rho$ and introduced $\cO_a(t) \equiv \ee^{t \cL^*} \cO_a$.
Here, $\cL^*$ is the adjoint Liouvillian, defined by $\Tr \, (\cL \rho) \cO = \Tr \rho \, \cL^* \cO$ for bounded operators $\cO$, yielding
$\cL^* \cO = \ii [H, \cO] + \sum_{j} \bigl( V_j^\dagger \cO V_j\pdag - \frac{1}{2}\{ V_j^\dagger V_j\pdag, \cO \} \bigr)$ \cite{Lindblad:1976}.
Since $\cL$ preserves Hermiticity, so does $\cL^*$, thus $\cO_a(t)^\dagger = \cO_a(t)$, and since $\cO_a$ has zero average, so does $\cO_a(t)$, due to that $\Tr \rho \, \cO_a(t) = \Tr \, (\Phi_t \rho) \cO_a = \Tr \rho \, \cO_a$.
Defining $\kappa_{ab}(t) \equiv \kappa_{ab}(t,0)$, we thus obtain Eqs.~\eqref{kappa_ab_result} and \eqref{kappa_ab_phi} unchanged apart from the unperturbed evolution in $\cO_a(t)$.
Note that the steps above and the final results are well-defined also in the open case for both finite- and infinite-dimensional Hilbert spaces (with the conditions on the observables $\cO_a$ in Part~\hyperref[Part:A3]{\ztitleref{Part:A3}}).

The main change is that the lengths of the tangent vectors $\XO_a(t)$ and $\KXO_a(t)$ in Eq.~\eqref{XaYa_t} can shrink.
Indeed, the lengths $\ell_{\rho}[\XO_a(t)] \equiv \sqrt{\cG_{\rho}( \XO_a(t), \XO_a(t))}$, measured by the Bures metric \eqref{cG_XY}, are just $\sqrt{\Tr \rho \, \cO_a(t)^2}$.
This is a nonincreasing function of $t$.
To prove this, note the semigroup property $\Phi_{t+s} = \Phi_{t} \Phi_{s} = \Phi_{s} \Phi_{t}$ ($t, s \geq 0$), and similarly $\Phi_{t+s}^* = \Phi_{s}^* \Phi_{t}^* = \Phi_{t}^* \Phi_{s}^*$ for the adjoint, which is $\Phi_{t}^* = \ee^{t\cL^*}$ for Lindblad dynamics.
Using the Kadison-Schwarz inequality \cite{Kadison:1952}, similar to the proof of Lemma~\hyperref[Lemma:3]{\ztitleref{Lemma:3}} in Part~\hyperref[Part:A3]{\ztitleref{Part:A3}} but for the operators $[\Phi_t^* (\Phi_s^*\cO_{\Lambda_n})] \sqrt{\rho}$, it follows that $\Tr \rho \, \cO_a(t+s)^2 \leq \Tr \rho \, \cO_a(s)^2$, from which the claim follows.
Moreover, by the same computation as in Eq.~\eqref{GX_X_ineq}, we have $\ell_{\rho}[\KXO_a(t)]^2
= \Tr \rho \, G_{K_\rho \XO_a(t)}^2
= \Tr \rho \, G_{\XO_a(t)}^2 - \sum_{j,k} \frac{2\prho_j\prho_k}{(\prho_j + \prho_k)^3} \bigl| \bigl( \XO_a(t) \bigr)_{jk} \bigr|^2 \leq \Tr \rho \, G_{\XO_a(t)}^2 = \ell_{\rho}[\XO_a(t)]^2$, with equality for all $\cO_a(t)$ if and only if $\rho$ is pure.
In conclusion,
\begin{equation}
\ell_{\rho}[\KXO_a(t)]
\leq \ell_{\rho}[\XO_a(t)]
\leq \ell_{\rho}[\XO_a(s)]
\end{equation}
for all $t \geq s \geq 0$.
This says that the (nonstrict) decay of $\XO_a(t)$'s length forms a (nonstrictly) shrinking envelope for $\KXO_a(t)$'s length, and the latter lies on the envelope for all $\cO_a(t)$ if and only if $\rho$ is pure.

%==========================================================
\vspace{0.0mm}\noindent\textbf{B.3. Transport}\vspace{0.5mm}\\
\csname phantomsection\endcsname%
\addcontentsline{toc}{subsection}{B.3. Transport}%
\label{Part:B3}%
\makeatletter\def\zref@titleref@current{B.3}\makeatother
\zlabel{Part:B3}%
%==========================================================
%
We work in $d$+1 spacetime dimensions and, for concreteness, consider a continuous system.
The case of, e.g., a $d$-dimensional lattice is covered by replacing integrals and derivatives by their discretizations.
We restrict to closed systems, as usual with a Hamiltonian $H$ that has at least one stationary $\rho \in \cD$. 

Let $Q_m$ labeled by $m = 1, \ldots, \cN$ be Hermitian operators that denote the system's conserved quantities, i.e., $[Q_m, H] = 0$.
At least one exists: The total energy $Q_2 = H$ (reserving $Q_1$ for the total number of particles or the total charge).
We assume that our $Q_m$ are sufficiently local, cf.\ \cite{Doyon:2017}, and write $Q_m = \int \dd^dx\, q_m(\bx)$ with densities $q_m(\bx)$ and introduce $\bjmath_m(\bx) = \bigl( \jmath_m^1(\bx), \ldots, \jmath_m^d(\bx) \bigr)$ as the corresponding currents satisfying the continuity equations
\begin{equation}
\label{cont_eq}
\partial_t q_m + \nabla \cdot \bjmath_m = 0.
\end{equation}
All the $q_m(\bx)$ and $\bjmath_m(\bx)$ are Hermitian operators, as are the total currents $\bJ_m = \int \dd^dx\, \bjmath_m(\bx)$.

Consider an initial stationary state $\rho$ that evolves under the dynamics \eqref{H_lambda} with perturbations, e.g., like $W_n = \int \dd^dy\, w_n(\by) q_n(\by)$. 
Here, $w_n(\by)$ are real-valued functions that produce imbalances in the system.
Basically, they are inhomogeneous thermodynamic fields, such as a position-dependent chemical potential or inverse temperature \cite{Spohn:1991, Moosavi:iCFT:2024, GLM:2018}.
One can then think of $\bJ_m = \bigl( J_m^1, \ldots, J_m^d \bigr)$ as observables, whose responses to the perturbations $W_n$ we seek.
However, one may as well perturb by $\bJ_n$, or measure responses in $W_m$; the latter correspond to total net currents that have flown through regions with nonzero gradients in $w_m(x)$.\footnote{$\partial_t W_m = \int \dd^dx\, w_m(x) \partial_t q_m(\bx) = - \dd^dx\, w_m(x) \nabla \cdot \bjmath_m(\bx) = \dd^dx\, \nabla w_m(x) \cdot \bjmath_m(\bx) + \text{boundary term}$.}
Concretely, the observables $\cO_a$ in this case are $\cO_m^{\mu}$, labeled by $a = (m, \mu)$ for $m = 1, \ldots, \cN$ and $\mu = 0, \ldots, d$, given by
\begin{subequations}
\label{cOs_def}
\begin{align}
\cO_{m}^{0}
& \equiv W_m - \Tr \rho \, W_m, \\
\cO_{m}^{i}
& \equiv J_m^{i} - \Tr \rho \, J_m^{i} \quad (i = 1, \ldots, d).
\end{align}
\end{subequations}
The functions $\lambda_b(s) = \lambda_n^\nu(s)$ in Eq.~\eqref{H_lambda} correspond to overall coupling strengths.
By definition, these $\cO_a$ are Hermitian and $\Tr \rho \, \cO_a = 0$.
Note that, if additionally $\Tr \rho \, \bJ_m = 0$ for all $m$, then we would say that $\rho$ is not only stationary but an equilibrium state.

The above is suitable, e.g., for quantum quenches where one may view the thermodynamic fields $w_n(\by)$ as given initial profiles, as in \cite{Moosavi:iCFT:2024, GLM:2018}.
However, one may also consider position-dependent responses and study observables $\cO_a$ with $a = (m, \mu, \bx)$ given by
\begin{equation}
\label{cOs_def_local}
\cO_{m}^{\mu}(\bx)
\equiv \jmath_m^{\mu}(\bx) - \Tr \rho \, \jmath_m^{\mu}(\bx) \quad (\mu = 0, 1, \ldots, d)
\end{equation}
with $\jmath_m^{0}(\bx) \equiv q_m(\bx)$.
In this case, the Hamiltonian \eqref{H_lambda} reads $H(s)
\equiv H - \sum_{n, \nu} \int \dd^d y\, \lambda_{n}^{\nu}(\by; s) \cO_{n}^{\nu}(\by)$ with the functions $\lambda_{b}(s) \equiv \lambda_{n}^{\nu}(\by, s)$ being the
thermodynamic fields themselves.
The responses $\kappa_{ab}(t)$ in Eqs.~\eqref{kappa_ab_result} and \eqref{kappa_ab_phi} then become
\begin{equation}
\label{jj_response}
\kappa_{m,n}^{\mu,\nu}(\bx, \by; t)
= -\ii \Tr \rho \bigl[ \cO_m^\mu(\bx, t), \cO_n^\nu(\by) \bigr]
\end{equation}
for $t > 0$.
One may also Fourier transform in $\bx$ and/or $\by$ to obtain
\begin{equation}
\label{jj_response_Fourier}
\kappa_{m,n}^{\mu,\nu}(\bp, \bq; t)
= \iint \dd^d x\, \dd^d y\, \kappa_{m,n}^{\mu,\nu}(\bx, \by; t) \, \ee^{\ii (\bp \cdot \bx + \bq \cdot \by)}
\end{equation}
for $t > 0$, in which case $a = (m, \mu, \bp)$ and $b = (n, \nu, \bq)$ are the labels, with $\bp = 0$ corresponding to studying the response in the total observable $\cO_{m}^{\mu}(\bp = 0) = \int \dd^d x\, \cO_{m}^{\mu}(\bx)$.
An example comes from determining whether a ground state is insulating, metallic, or superconducting using current-current correlations \cite{ScalapinoWhiteZhang:1993}.
Specifically, one considers the susceptibilities $\kappa_{1,1}^{i,j}(\bx, \by; t)$, in which case $\lambda_{1}^{j}(\by, s)$ are the components $A^j(\by,s)$ of an external vector potential, and each associated Drude weight [cf.\ Eq.~\eqref{kappa_gen_grounds}] is a superfluid weight \cite{PeottaTorma:2015, YuEtAl:2025}.

We can also include other sets of operators, such as the (generalized) \emph{polarization density} $\bP(\bx) = \bigl( P^1(\bx), \ldots, P^d(\bx) \bigr)$ and its corresponding \emph{polarization current} $\bjmath_{\textrm{P}}(\bx) = \bigl( \jmath_{\textrm{P}}^1(\bx), \ldots, \jmath_{\textrm{P}}^d(\bx) \bigr)$, where each $\jmath_{\textrm{P}}^{i}(\bx) = \partial_{t} P^{i}(\bx)$.
One may consider the corresponding charge density $q_{\textrm{P}}(\bx) = - \nabla \cdot \bP(\bx)$, but generally one is interested in responses to perturbations in the individual fields $\lambda_{\textrm{P},j,\by}(s) = \lambda_{\textrm{P}}^j(\by,s)$, each conjugate to a $P^j(\by)$.
The observables are\footnote{The sign for $\cO_{\textrm{P},i, \bx}$ is consistent with Eqs.~\eqref{kappa_ab_def}--\eqref{kappa_ab_phi}, since the conjugate field $\boldsymbol{\lambda}_{\textrm{P}}(\bx) = - \nabla \lambda_{\textrm{P}}^0(\bx)$ enters the perturbed Hamiltonian \eqref{H_lambda} with the opposite sign compared to the potential $\lambda_{\textrm{P}}^0(\bx)$ associated to  $q_{\textrm{P}}(\bx)$:
$\int \dd^d y\, \lambda_{\textrm{P}}^0(\by,s) q_{\textrm{P}}(\by)
= - \int \dd^d y\, \lambda_{\textrm{P}}^0(\by,s) \nabla \cdot \bP(\by)
= - \int \dd^d y\, \boldsymbol{\lambda}_{\textrm{P}}(\by,s) \cdot \bP(\by) + \text{boundary}\\ \text{term}$.}
\begin{subequations}
\label{cOs_def_local_pd}
\begin{align}
\cO_{\textrm{P}}^{i}(\bx)
& \equiv - P^{i}(\bx) + \Tr \rho \, P^{i}(\bx) \quad (i = 1, \ldots, d), \\
\cO_{\textrm{J}}^{i}(\bx)
& \equiv \jmath_{\textrm{P}}^{i}(\bx) - \Tr \rho \, \jmath_{\textrm{P}}^{i}(\bx) \quad (i = 1, \ldots, d),
\end{align}
\end{subequations}
and the responses in Eq.~\eqref{jj_response} are complemented by
\begin{subequations}
\begin{align}
\kappa_{\textrm{P},\textrm{P}}^{i,j}(\bx, \by; t)
& = - \ii \Tr \rho \bigl[ \cO_{\textrm{P}}^i(\bx, t), \cO_{\textrm{P}}^j(\by) \bigr], \label{electrical_susceptibility} \\
\kappa_{\textrm{J},\textrm{P}}^{i,j}(\bx, \by; t)
& = \ii \partial_t \Tr \rho \bigl[ \cO_{\textrm{P}}^i(\bx, t), \cO_{\textrm{P}}^j(\by) \bigr], \label{optical_conductivity}
\end{align}
\end{subequations}
and so on, for $t > 0$.
(One may also Fourier transform these in $\bx$ and/or $\by$.)
An example comes from bound charges in electrodynamics, in which case the $\lambda_{\textrm{P}}^j(\by,s)$ are the components $E^j(\by,s)$ of an external electric field, $\chi_{\textrm{e}}^{i,j}(\bx, \by; t) \equiv - \kappa_{\textrm{P},\textrm{P}}^{i,j}(\bx, \by; t)$ is the \emph{electrical susceptibility}, and $\sigma_{i,j}(\bx, \by; t) \equiv \kappa_{\textrm{J},\textrm{P}}^{i,j}(\bx, \by; t) = \partial_t \chi_{\textrm{e}}^{i,j}(\bx, \by; t)$ is the \emph{optical conductivity}.

%==========================================================
\vspace{0.0mm}\noindent\textbf{B.4. Conductivities and susceptibilities}\vspace{0.5mm}\\
\csname phantomsection\endcsname%
\addcontentsline{toc}{subsection}{B.4. Conductivities and susceptibilities}%
\label{Part:B4}%
\makeatletter\def\zref@titleref@current{B.4}\makeatother
\zlabel{Part:B4}%
%==========================================================
%
Whether response functions $\kappa_{ab}(t)$ in Eqs.~\eqref{kappa_ab_result} and~\eqref{kappa_ab_phi} are conductivities or susceptibilities depends on the observables.
A conductivity measures the response in a current to a perturbation in the field conjugate to a density, so, e.g., $a = (m, i, \bp)$, $b = (n, 0, \bq)$ in the notation above.
On the other hand, a susceptibility measures the response in an observable to a perturbation in the field conjugate to the same type of observable, so, e.g., $a = (m, i, \bp)$, $b = (n, i, \bq)$ (both currents) or $a = (m, 0, \bp)$, $b = (n, 0, \bq)$ (both densities).
Thus, using the continuity equation \eqref{cont_eq} [or $\jmath_{\textrm{P}}^{i}(\bx) = \partial_{t} P^{i}(\bx)$ for polarization], a conductivity basically corresponds to the time derivative of some susceptibility.

Let $\kappa_{ab}(t)$ be a conductivity for a total current, say, the response \eqref{jj_response_Fourier} for $a = (m, i, \bp = 0)$ and $b = (n, 0, \bq)$.
Then, on general grounds,
\begin{equation}
\label{kappa_gen_grounds}
\kappa_{ab}(t) = D_{ab} + \kappa_{ab}^{\textrm{reg}}(t)
\end{equation}
for $t > 0$ with constants $D_{ab}$ and functions $\kappa_{ab}^{\textrm{reg}}(t)$ decaying with time $t$; see, e.g., \cite{Moosavi:iCFT:2024, GLM:2018, Spohn:2018}.
For a closed system, $\kappa_{ab}(t)$ in Eq.~\eqref{kappa_gen_grounds} is understood as the response in the infinite-volume limit for $t$ finite, since its dynamics is typically almost periodic in finite volume.
Each $D_{ab}$ is a Drude weight, which if nonzero indicates ballistic transport, and each $\kappa_{ab}^{\textrm{reg}}(t)$ is the associated regular part, which if nonzero indicates (anomalous) diffusive contributions.
This is usually phrased using their Fourier transforms [cf.\ Eq.~\eqref{kappa_ab_phi}]
\begin{subequations}
\begin{equation}
\label{kappaF}
\chi_{ab}(\omega)
\equiv
\int_{-\infty}^{\infty} \dd t\, \kappa_{ab}(t) \, \ee^{-\ii \omega t} = \int_{0}^{\infty} \dd t\, \phi_{ab}(t) \, \ee^{-\ii \omega t}
\end{equation}
for frequencies $\omega \in \mathbb{R}$:
\begin{equation}
\label{kappaF_Dreg}
\chi_{ab}(\omega)
= D_{ab} \Bigl[ \pi \delta(\omega) - \operatorname{p.\!v.\!} \frac{\ii}{\omega} \Bigr] + \chi^{\textrm{reg}}_{ab}(\omega),
\end{equation}
where $\operatorname{p.\!v.\!}$ denotes principal value, or rather in terms of their real parts,\footnote{Since the real part of the symmetrized conductivity tensor in frequency space is dissipative; cf.\ Part~\hyperref[Part:B5]{\ztitleref{Part:B5}}.}
\begin{equation}
\label{kappaF_Dreg_Re}
\Re \chi_{ab}(\omega)
= D_{ab} \pi \delta(\omega) + \Re \chi^{\textrm{reg}}_{ab}(\omega).
\end{equation}
\end{subequations}
The $\omega = 0$ value of $\Re \chi^{\textrm{reg}}_{ab}(\omega)$ gives the elements of the Onsager matrix $(L_{ab})$.
If a certain $L_{ab}$ is nonzero and finite, then that $(a,b)$ component has a normal diffusive contribution.
Other diffusive contributions are also possible:
Superdiffusion corresponds to a diverging $\Re \chi^{\textrm{reg}}_{ab}(\omega)$ as $\omega \to 0$, and subdiffusion to a $\Re \chi_{ab}^{\textrm{reg}}(\omega)$ that vanishes as a nonanalytic power law.\footnote{For concreteness, consider the following representative terms:
$\kappa_{ab}^{\textrm{reg}}(t) = A \ee^{-t/\tau} + B \cos(\pi\gamma/2) t^{-\gamma}$, where $A, B \geq 0$, $\tau > 0$, and $2 > \gamma > 0$.
It follows that $\Re \chi_{ab}^{\textrm{reg}}(\omega) =
\frac{A \tau}{1 + \omega^2 \tau^2} + \frac{B\pi}{2\Gamma(\gamma)} |\omega|^{\gamma-1}$, where $\frac{A \tau}{1 + \omega^2 \tau^2} = A \tau + O(\omega^2)$.
Here, the term $\propto A$ corresponds to normal diffusion, and the term $\propto B$ to superdiffusion ($1 > \gamma > 0$) or subdiffusion ($2 > \gamma > 1$).}

%==========================================================
\vspace{0.0mm}\noindent\textbf{B.5. Sum rules}\vspace{0.5mm}\\
\csname phantomsection\endcsname%
\addcontentsline{toc}{subsection}{B.5. Sum rules}%
\label{Part:B5}%
\makeatletter\def\zref@titleref@current{B.5}\makeatother
\zlabel{Part:B5}%
%==========================================================
%
Recall from Eq.~\eqref{kappa_ab_phi} that $\kappa(t) = \theta(t) \phi(t)$
with $\phi(t)$ of the form $-\ii \Tr \rho \bigl[ \cO_a(t), \cO_b \bigr]$.
Its Fourier transform [cf.\ Eq.~\eqref{kappaF}] splits into real and imaginary parts as $\chi(\omega) = \chi'(\omega) - \ii \chi''(\omega)$.
Sum rules can then be viewed as relations for the moments of the frequency distribution of the dissipative part \cite{Kubo:1966}.
I.e., integrals like
\begin{equation}
\label{sum_rule_integrals}
\int_{0}^{\infty} \dd\omega\, \omega^{n} D(\omega)
\end{equation}
for $n \in \mathbb{Z}$ with $D(\omega)$ either $\chi'(\omega) \equiv \Re \chi(\omega)$ or $\chi''(\omega) \equiv -\Im  \chi(\omega)$, whichever is dissipative.

In short, $\chi'$ is dissipative if $\phi(t)$ is even in $t$, and $\chi''$ is dissipative if it is odd.
The reason is that the inverse Fourier transform becomes $\phi(t) = \frac{1}{\pi} \int_{-\infty}^{\infty} \dd \omega\, \chi'(\omega) \ee^{\ii \omega t}$ or 
$\phi(t) = \frac{1}{\pi \ii} \int_{-\infty}^{\infty} \dd \omega\, \chi''(\omega) \ee^{\ii \omega t}$, respectively, thus $\chi''$ ($\chi'$) does not contribute in the former (latter) case (see Secs.~7 and 10 in \cite{Kubo:1966}).
Consequently, when computing the absorbed energy 
$\int_{t_0}^{t} \! \dd s \Tr \rho(s) \, \partial_s H(s) = \underset{a, b}{\sum} \int_{t_0}^{t} \! \dd s \bigl[ \lambda_a(t) \lambda_b(s) \kappa_{ab}(t-s) - \int_{t_0}^s \! \dd u\, \lambda_a(s) \lambda_b(u) \partial_s \kappa_{ab}(s-u) \bigr] + O(\lambda^3)$ (cf.\ Parts~\hyperref[Part:B1]{\ztitleref{Part:B1}} and \hyperref[Part:B2]{\ztitleref{Part:B2}}), then only dissipative parts can contribute.
The noncontributing parts are correspondingly said to be nondissipative (or reactive).
For instance, if $\kappa_{ab}(t)$ is a susceptibility tensor (see Part~\hyperref[Part:B4]{\ztitleref{Part:B4}}), then the symmetric part of $\phi_{ab}(t)$ is odd in $t$ and so its imaginary part in frequency space is dissipative, while the antisymmetric part of $\phi_{ab}(t)$ is even in $t$ and so its real part in frequency space is dissipative.
Since conductivities correspond to the time derivative of susceptibilities (again see Part~\hyperref[Part:B4]{\ztitleref{Part:B4}}), the roles of which part is even or odd under changing $t \to -t$ and which part is dissipative or nondissipative are then reversed.

As an example, we derive sum rules for the $\omega^{-1}$-moment of the \emph{generalized optical conductivities} $\ii \partial_t \Tr \rho \bigl[ \cO_a(t), \cO_b \bigr]$ [cf.\ Eq.~\eqref{optical_conductivity}].
We show that one can express them in terms of the Bures metric by using our fluctuation-dissipation structure, i.e., by going via the right then top relation in Fig.~\ref{Fig:Connections}.
Indeed, due to Eq.~\eqref{ashaped_eq}, this holds in general, for closed and open systems alike, while below we will restrict to unitary dynamics.
Our approach is different from \cite{SWM:2000}, which derived the original SWM sum rule for insulators in their ground states by going via polarization fluctuations \cite{RestaVanderbilt:2007}, i.e., the bottom then left relation in Fig.~\ref{Fig:Connections} pulled back to a parameter space (cf.\ Part~\hyperref[Part:B6]{\ztitleref{Part:B6}} below).
We will derive our sum rules without going into any many-body details.
In fact, all we will use is that we compute the response in an observable $-\partial_t \cO_a$ to a perturbation in the field conjugate to $\cO_b$ for a closed system in any state $\rho \in \cD$ that is stationary under the Hamiltonian $H$:
\begin{equation}
\sigma_{ab}(t) = \theta(t) \varphi_{ab}(t),
\quad
\varphi_{ab}(t) = \ii \partial_t \Tr \rho \, [\cO_a(t), \cO_b].
\end{equation}
Using Eq.~\eqref{ashaped_eq}, we immediately obtain
\begin{equation}
\label{kappa_ab_SWM}
\varphi_{ab}(t)
= - \partial_t \cF_{\rho}(X_a(t), X_b)
= - 2 \partial_t \cG_{\rho}(Y_a(t), X_b),
\end{equation}
where $Y_a(t) = \ee^{\ii Ht} K_{\rho} X_a \ee^{-\ii Ht}$ and $X_a = \cO_a \rho + \rho \cO_a$.
This already contains the essence of all sum rules below, and for all other moments.
Indeed, Eq.~\eqref{kappa_ab_SWM} and more generally Eq.~\eqref{ashaped_eq} say that any linear response and thus any moment can be expressed entirely and equivalently in terms of the metric or the curvature.

Since $\varphi_{ab}(t) = \varphi_{ba}(-t)$, consider instead the symmetrized $\varphi_{ab}^{\textrm{S}}(t) \equiv \frac{1}{2} \bigl[ \varphi_{ab}(t) + \varphi_{ba}(t) \bigr]$ and the antisymmetrized $\varphi_{ab}^{\textrm{A}}(t) \equiv \frac{1}{2} \bigl[ \varphi_{ab}(t) - \varphi_{ba}(t) \bigr]$, which satisfy $\varphi_{ab}^{\textrm{S}}(-t) = \varphi_{ab}^{\textrm{S}}(t)$ and $\varphi_{ab}^{\textrm{A}}(-t) = -\varphi_{ab}^{\textrm{A}}(t)$.
By the recipe above, $\chi_{ab}^{\textrm{S}}{}'(\omega) = \Re \int_0^\infty \dd t\, \varphi_{ab}^{\textrm{S}}(t) \ee^{-\ii \omega t}$ and $\chi_{ab}^{\textrm{A}}{}''(\omega) = - \Im \int_0^\infty \dd t\, \varphi_{ab}^{\textrm{A}}(t) \ee^{-\ii \omega t}$ are dissipative.
We only study the former here, while the latter and the nondissipative parts can be studied in the same way.

\noindent
{\bf Pure states.}
First, consider a pure state $\rho_\psi \equiv |\psi\rangle\langle\psi|$ for some nonzero $|\psi\rangle \in \cH$ ($\langle\psi|\psi\rangle = 1$).
Since $|\psi\rangle$ is stationary, i.e., $[\rho_\psi, H] = 0$, it must be an eigenstate of $H$.
Note that $H$ only has to have a single eigenstate, namely $|\psi\rangle$.
Without loss of generality, let $|\psi\rangle = |e_1\rangle$.
Then, by the spectral theorem, we have an orthogonal decomposition $H = E_1 |e_1\rangle\langle e_1| + \tilde{H}$, where $\tilde{H}$ is general and could have only a continuous spectrum.

What if $\tilde{H}$ has only a point spectrum?
In this case, we can denote the other eigenstates of $H$ as $|e_k\rangle$ for $k \geq 2$.
Eqs.~\eqref{Ya_t}, \eqref{cGcF_XY_pure_jk}--\eqref{Krho_pure_jk}, and \eqref{kappa_ab_SWM} then imply that%
\begin{subequations}
\begin{multline}
\label{dt_Ya_t_psi}
\partial_t Y_a(t)
= \sum_{k \geq 2} \Bigl[ X^a_{1k} |e_1\rangle\langle e_k| \omega_{1k} \, \ee^{\ii\omega_{1k}t} \\
- X^a_{k1} \omega_{k1} \, \ee^{\ii\omega_{k1}t} |e_k\rangle\langle e_1| \Bigr]
\end{multline}
and
\begin{multline}
\varphi_{ab}(t)
= - \sum_{k \geq 2} \Bigl [X^a_{1k} X^b_{k1} \omega_{1k} \, \ee^{\ii\omega_{1k}t} \\
- X^b_{1k} X^a_{k1} \omega_{k1} \, \ee^{\ii\omega_{k1}t} \Bigr],
\end{multline}
\end{subequations}
using the differences  $\omega_{jk} = E_j - E_k$ between eigenenergies of $H$.
By symmetrizing, Fourier transforming, and computing the frequency integral, we obtain
\begin{equation}
\int_0^\infty \dd\omega\, \frac{\chi_{ab}^{\textrm{S}}{}'(\omega)}{\omega}
= \pi \cG_{\rho_{\psi}}(X_a^+ - X_a^-,X_b) \label{SWM_sum_rule_statpure}
\end{equation}
for any stationary pure state $|\psi\rangle = |e_1\rangle$ and $\cG_{\rho_\psi}$ as in Eq.~\eqref{cG_XY_pure_jk}.
Here, $X_a^+$ contains the elements $X_{k1}$, $X_{1k}$ for which $\omega_{k1} > 0$, while $X_a^-$ contains those for which $\omega_{k1} < 0$.
Elements corresponding to $\omega_{k1} = 0$ cancel from the derivation.
There are two extremal cases:\vspace{-1.5mm}
\begin{enumerate}[topsep=0.25ex, itemsep=-0ex, partopsep=0.25ex, parsep=0.75ex, leftmargin=4.00ex]

\item
If $H$ is bounded from below and $|e_1\rangle$ is the ground state, then $X_a^- = 0$.
If it is also nondegenerate, then $X_a^+ = X_a$ and we obtain Eq.~\eqref{SWM_sum_rule_GS} and $\int_0^\infty \dd\omega\, \omega^{-1} \chi_{aa}^{\textrm{S}}{}'(\omega) > 0$.

\item
If $H$ is bounded from above, we can let $|\psi\rangle = |e_1\rangle$ be the highest-energy state.
If so, then $X_a^+ = 0$.
If that state is also nondegenerate, then $X_a^- = X_a$ and Eq.~\eqref{SWM_sum_rule_statpure} implies that $\int_0^\infty \dd\omega\, \omega^{-1} \chi_{aa}^{\textrm{S}}{}'(\omega) < 0$.\vspace{-1mm}

\end{enumerate}

Next, let us see the general case where $\tilde{H}$ can have a nonempty continuous spectrum.
There are many ways to proceed, including using spectral measures.
Here we will instead use functional calculus and formulas similar to above by noting that there is still a basis $|e_k\rangle$ ($k \geq 2$) for the subspace $\ker \rho_\psi$ of $\cH$ on which $\tilde{H}$ is defined, since the Hilbert space is separable.
However, these states are no longer guaranteed to be eigenstates of $\tilde{H}$.
By the same arguments as before, we again obtain Eq.~\eqref{dt_Ya_t_psi} but with $\omega_{1k} = -\omega_{k1}$ replaced by operators $\omega_{1\tilde{H}} \equiv E_1 - \tilde{H} \equiv - \omega_{\tilde{H}1}$, so $\varphi_{ab}(t)$ can contain scattering events between the states:
\begin{multline}
\varphi_{ab}(t)
= - \sum_{k,k' \geq 2} \Bigl [X^a_{1k} X^b_{k'1} \langle e_{k}| \omega_{1\tilde{H}} \, \ee^{\ii\omega_{1\tilde{H}}t} |e_{k'}\rangle \\
- X^b_{1k'} X^a_{k1} \langle e_{k'}| \omega_{\tilde{H}1} \, \ee^{\ii\omega_{\tilde{H}1}t} |e_{k}\rangle \Bigr].
\end{multline}
Proceeding similar to before, using functional calculus for the Fourier transform and the frequency integral, we get
\begin{equation}
\int_0^\infty \dd\omega\, \frac{\chi_{ab}^{\textrm{S}}{}'(\omega)}{\omega}
= \pi \cG_{\rho_\psi}(VX_a, X_b)
\label{SWM_sum_rule_statpure_scattering}
\end{equation}
for $V \equiv \theta(\tilde{H} - E_1) - \theta(E_1 - \tilde{H})$ acting in $\ker \rho_\psi$ and $\cG_{\rho_\psi}$ again as in Eq.~\eqref{cG_XY_pure_jk}.
This is precisely Eq.~\eqref{SWM_sum_rule_statpure} if we define $X_a^+ \equiv \theta(\tilde{H} - E_1) X_a$ and $X_a^- \equiv \theta(E_1 - \tilde{H}) X_a$.
Thus, independent of the nature of the spectrum $\sigma_{\tilde{H}}$ of $\tilde{H}$, we have the same extremal cases:\vspace{-1.5mm}
\begin{enumerate}[topsep=0.25ex, itemsep=-0ex, partopsep=0.25ex, parsep=0.75ex, leftmargin=4.00ex]

\item
If $H$ is bounded from below, and if $|e_1\rangle$ is nondegenerate and lies below or at the bottom of $\sigma_{\tilde{H}}$, then Eq.~\eqref{SWM_sum_rule_statpure_scattering} becomes Eq.~\eqref{SWM_sum_rule_GS}.

\item
If $H$ is bounded from above, and if $|e_1\rangle$ is nondegenerate and lies above or at the top of $\sigma_{\tilde{H}}$, then the right-hand side of Eq.~\eqref{SWM_sum_rule_statpure_scattering} is $- \pi \cG_{\rho_\psi}(X_a, X_b)$.\vspace{-1.5mm}

\end{enumerate}
The above shows that Eq.~\eqref{SWM_sum_rule_GS} is valid for any Hamiltonian $H$ with a nondegenerate ground state, regardless of the nature of the rest of its spectrum.
In particular, pulling the geometry back to a suitable parameter space for many-body electronic systems yields the original SWM sum rule \cite{SWM:2000}.
This also shows that the type of sum rule \eqref{SWM_sum_rule_GS} generalizes beyond the view in \cite{OnishiFu:2024} that limited their proposal to insulators.

\noindent
{\bf Mixed states.}
Second, consider any stationary $\rho \in \cD$.
For simplicity, assume that $H$ has only a point spectrum; Hamiltonians with a nonempty continuous spectrum can be handled as in the pure case.
Since $[\rho, H] = 0$, we can simultaneously diagonalize $\rho = \sum_{j} \prho_j |e_j\rangle \langle e_j|$ and $H = \sum_{j} E_j |e_j\rangle \langle e_j|$.
Eqs.~\eqref{Ya_t}, \eqref{cGcF_XY_jk}--\eqref{Krho_jk}, and \eqref{kappa_ab_SWM} then imply that
\begin{subequations}
\begin{equation}
\label{dt_Ya_t_rho}
\partial_t Y_a(t)
= \sum_{j,k} \frac{\prho_j - \prho_k}{\prho_j + \prho_k} X^a_{jk} \omega_{jk} \, \ee^{\ii\omega_{jk}t} |e_j\rangle\langle e_k| \vspace{-1mm}
\end{equation}
and
\begin{equation}
\varphi_{ab}(t)
= - \sum_{j,k}\frac{\prho_j - \prho_k}{(\prho_j + \prho_k)^2} \overline{X^a_{jk}} X^b_{jk} \omega_{jk} \, \ee^{-\ii\omega_{jk}t}.
\end{equation}
\end{subequations}
Proceeding exactly as in the pure case, we obtain
\begin{equation}
\label{SWM_sum_rule_statmix}
\int_0^\infty \dd\omega\, \frac{\chi_{ab}^{\textrm{S}}{}'(\omega)}{\omega}
= \frac{\pi}{2} \cF_{\rho}(X_a^{H},X_b)
= \pi \cG_{\rho}(Y_a^{H},X_b),
\end{equation}
expressed using either the Uhlmann curvature $\cF_\rho$ or the Bures metric $\cG_\rho$ for vectors $X_a^{H}$ and $Y_a^{H} \equiv K_\rho X_a^{H}$ with elements $X^{a,H}_{jk} = - \ii \sgn(\omega_{jk}) X^a_{jk}$ and $Y^{a,H}_{jk} = - \sgn(\omega_{jk}) \frac{\prho_j - \prho_k}{\prho_j + \prho_k} X^a_{jk}$, respectively.

\noindent
{\bf Thermal states.}
As a special case, given a Hamiltonian $H$ such that $Z_\beta \equiv \Tr \ee^{-\beta H} < \infty$ for $\beta > 0$, consider $\rho_\beta \equiv Z_{\beta}^{-1} \ee^{-\beta H}$.
Then $\prho_j = Z_{\beta}^{-1} \ee^{-\beta E_j}$ and we get Eq.~\eqref{SWM_sum_rule_statmix} for $Y_a^{H} = K_{\rho_\beta} X_a^{H}$ with elements $Y^{a,H}_{jk} = \tanh(\beta |\omega_{jk}|/2) X^a_{jk}$.
[This case could have been obtained more directly using Eq.~\eqref{Krho_beta_jk}.]
In the zero-temperature limit, $\lim_{\beta \to \infty} K_{\rho_\beta} X_a^{H} = X_a^+$, which gives Eq.~\eqref{SWM_sum_rule_statpure} with $X_a^- = 0$, and Eq.~\eqref{SWM_sum_rule_GS} if the ground state is nondegenerate.
The version in \cite{HHTZ:2016, JiEtAl:2025} with an extra factor $\tanh(\beta\omega/2)$ in the frequency integral for the $\omega^{-1}$-moment corresponds to the same thermal sum rule as here, but expressed using the observables $\cO_a = G_{X_a}$ and not the tangent vectors $X_a$ [cf.\ Eqs.~\eqref{Xa_t} and \eqref{G_jk}].
Explicitly, by repeating the derivation above with the extra factor, we find
\begin{multline}
\int_0^\infty \dd\omega\, \tanh(\beta\omega/2) \frac{\chi_{ab}^{\textrm{S}}{}'(\omega)}{\omega} \\
= \pi \cG_{\rho_\beta}(K_{\rho_\beta}X_a, K_{\rho_\beta}X_b)
= \frac{\pi}{4} \cG^{\textrm{QFI}}_{\rho_\beta}(\cO_a, \cO_b),
\end{multline}
in agreement with \cite{HHTZ:2016}, where\vspace{-0.5mm}
\begin{equation}
\label{G_QFI}
\cG^{\textrm{QFI}}_\rho(\cO_a,\cO_b)
= 2 \sum_{j,k} \frac{(\prho_j - \prho_k)^2}{\prho_j + \prho_k} \overline{(\cO_a)_{jk}} (\cO_b)_{jk} \vspace{-0.5mm}
\end{equation}
is the (positive-definite symmetric) bilinear form corresponding to the quantum Fisher information $Q^{\textrm{QFI}}_\rho(\cO) \equiv \cG^{\textrm{QFI}}_\rho(\cO,\cO)$.
However, that approach does not extend to general mixed states, since it relies on $\prho_j = \ee^{-\beta E_j}$ to convert $\tanh(\beta\omega/2)$ to $\tanh(\beta\omega_{kj}/2) = (\prho_j - \prho_k)/(\prho_j + \prho_k)$ when integrating it against the delta function from the Fourier transform of $\varphi_{ab}(t)$.

Expressing this type of sum rules using the quantum Fisher information for general stationary density operators $\rho$ can be done by formally rewriting Eq.~\eqref{SWM_sum_rule_statmix} as
\begin{equation}
\int_0^\infty \dd\omega\, \frac{\chi_{ab}^{\textrm{S}}{}'(\omega)}{\omega}
= \frac{\pi}{4} \cG^{\textrm{QFI}}_\rho \bigl( G_{X_a^{H}}, G_{K_{\rho}^{-1} X_b} \bigr).
\end{equation}
Here, $G_{X_a^{H}}$ and $G_{K_{\rho}^{-1}X_b}$ are the respective solutions of Eq.~\eqref{Lyapunov_eq} for $X_a^{H}$ and $K_{\rho}^{-1}X_b$.
Note the appearance of the inverse $K_{\rho}^{-1}$, which need not exist, and $K_{\rho}^{-1}X$ is not guaranteed to be a tangent vector for all allowed $X$, except when $\rho$ is pure.
On the other hand, Eq.~\eqref{SWM_sum_rule_statmix} in terms of the Uhlmann curvature or the Bures metric holds in general, without any such qualifications.

%==========================================================
\vspace{0.0mm}\noindent\textbf{B.6. Pullback to parameter spaces}\vspace{0.5mm}\\
\csname phantomsection\endcsname%
\addcontentsline{toc}{subsection}{B.6. Pullback to parameter spaces}%
\label{Part:B6}%
\makeatletter\def\zref@titleref@current{B.6}\makeatother
\zlabel{Part:B6}%
%==========================================================
%
The picture on a parameter space $\cM$ arises under a pull back by a map $\cM \ni \alpha \mapsto \rho_\alpha \in \cD$, provided the metric remains nondegenerate.\footnote{If true, then the induced metric is Riemannian (since positive semidefinite and symmetric are automatic, and dependence on $\alpha$ is smooth as long as $\rho_\alpha$ depends smoothly on $\alpha$).}
That is all we need, with one qualification:
The observables $\cO_a$ in Eqs.~\eqref{kappa_ab_result}--\eqref{fluct_cG} are unaffected by the geometry being on $\cD$ (i.e., geometry of quantum states) or $\cM$ (i.e., quantum geometry) as long as they correspond to an allowed tangent vector to $\cM$.
If they do not, then one needs to consider a bigger parameter space, or constrain the choice of observables.
This is not really an issue since both the choice of parametrization and observables is motivated by the same physics.
However, it may explain why, to our knowledge, the conventional part in the superfluid weight \cite{PeottaTorma:2015, YuEtAl:2025} has so far not been recognized to also be geometric.

What changes is that the states belong to some family $\rho_\alpha$ labeled by $\alpha \in \cM$, and the relevant tangent vectors are those to $\cM$, on which the pullbacks $\cG_{\alpha}$ and $\cF_{\alpha}$ act.
The latter can then be used to define the action of the corresponding induced $K_{\alpha}$, just as in Eq.~\eqref{FDS}, if $\cG_{\alpha}$ is nondegenerate.\footnote{The tangent vectors to $\cM$ generally generate a subset of the tangent vectors to $\cD$. Thus, the induced $K_{\alpha}$ is not a pullback of $K_{\rho}$, but of $K_{\rho}$ projected to the given subset. This is automatic by defining $K_{\alpha}$ from the pullbacks $\cG_{\alpha}$ and $\cF_{\alpha}$.}
A typical example is $\rho_\alpha = U_\alpha \rho U_\alpha^{-1}$ for a given $\rho \in \cD$ and some unitary representation $U_\alpha$ of $\alpha \in \cM$.
In this case, the tangent vectors are $\dd\rho_\alpha = [u_{\bX}, \rho_\alpha] \equiv 
X_{\alpha}$, where $u_{\bX}$ is an anti-Hermitian representation of the tangent vector $\bX$ to $\cM$ at $\alpha \in \cM$ that maps to a tangent vector $X_\alpha$ to $\cD$ at $\rho_\alpha$.
It then follows that, e.g.,
$\cG_\alpha(\bX,\bY)
\equiv
\cG_{\rho_\alpha}(X_{\alpha},Y_{\alpha})
= \frac{1}{4} \cG^{\textrm{QFI}}_{\rho_\alpha}(u_{\bX}, u_{\bY})$
using the quantum Fisher information bilinear form in Eq.~\eqref{G_QFI}.

As mentioned in the main text, there is no practically useful parametrization for the whole $\cD$ in general, even when $\dim(\cH) = N < \infty$.
A typical choice is to parametrize using the generators of $\operatorname{su}(N)$ (see Ch.~8 in \cite{BeZy:2009}).
However, it then still remains to determine if a point lies in the parameter space, which is generally only known implicitly from a system of $N$ coupled equations, so only small $N$ are feasible.

%==========================================================
\vspace{0.0mm}\noindent\textbf{B.7. Beyond linear response}\vspace{0.5mm}\\
\csname phantomsection\endcsname%
\addcontentsline{toc}{subsection}{B.7. Beyond linear response}%
\label{Part:B7}%
\makeatletter\def\zref@titleref@current{B.7}\makeatother
\zlabel{Part:B7}%
%==========================================================
%
It would be interesting to extend the picture in Fig.~\ref{Fig:Connections} to nonlinear response.
For fluctuations, this should correspond to including higher-order correlations.
For the geometry, one may expect needing the expansion of the Bures distance beyond the Riemannian metric (the Bures metric) by including its Riemannian curvature and so on, and the expansion of the holonomy beyond the contribution from the Uhlmann curvature.

%==========================================================
\end{document}